%% file: rfcnlp_main.tex
\documentclass[10pt, conference, compsocconf]{IEEEtran}

\usepackage{times}
\usepackage{latexsym}
\usepackage{graphicx}
\usepackage{listingsutf8}
\usepackage{float}
\usepackage{multirow}
\usepackage[dvipsnames,table]{xcolor}
\usepackage{soul,xspace}
\usepackage{amsmath}
\usepackage{amssymb}
\usepackage{amsfonts}
\usepackage{amsthm}
\usepackage{bm}
\usepackage{tikz}
\usepackage{adjustbox}
\usepackage[pdf]{graphviz}
\usepackage{pifont}
\usepackage{enumerate}
\usepackage{subcaption}

\usepackage{tcolorbox}

\newcommand{\cmark}{\ding{51}}%
\newcommand{\xmark}{\ding{55}}%
\usetikzlibrary{shapes,arrows}

\usetikzlibrary{automata, positioning, arrows, calc}
    \tikzset{
        >=stealth, 
        node distance=3.5cm, 
        every state/.style={rectangle, thick, fill=gray!10}, 
        initial text=$ $, 
    }

\newcommand{\etal}{\textit{et~al}.}

\newif\ifRevisionDraftThree

\newif\ifRevisionDraftTwo

\newif\ifRevisionDraft

\usepackage{microtype}
\usepackage{comment}

\newcommand{\xmlrep}{intermediary representation}
\newcommand{\fsm}{FSM}
\newcommand{\program}{\textsc{Promela} program}

\newcommand{\bert}{\textsc{NeuralCRF}}
\newcommand{\linear}{\textsc{LinearCRF}}
\newcommand{\bertrules}{\textsc{NeuralCRF+R}}
\newcommand{\linearrules}{\textsc{LinearCRF+R}}

\newcommand{\xml}[1]{\lstinline[columns=fixed]{#1}}

\title{Automated Attack Synthesis by Extracting Finite State Machines from Protocol Specification Documents}

\author{\IEEEauthorblockN{Maria Leonor Pacheco\IEEEauthorrefmark{1},
  Max von Hippel\IEEEauthorrefmark{2},
  Ben Weintraub\IEEEauthorrefmark{2},
  Dan Goldwasser\IEEEauthorrefmark{1},
  Cristina Nita-Rotaru\IEEEauthorrefmark{2}}

\IEEEauthorblockA{\IEEEauthorrefmark{1}Purdue University, West Lafeyette, IN, USA,
  \{pachecog,dgoldwas\}@purdue.edu}

\IEEEauthorblockA{\IEEEauthorrefmark{2}Northeastern University, Boston, MA, USA,
  \{vonhippel.m,weintraub.b,c.nitarotaru\}@northeastern.edu}}

\begin{document}
\maketitle

\input{abstract}

\input{intro}
\input{bk}
\input{annotation}

\input{embedding}

\input{parser}

\input{translation}

\input{task}

\input{evaluation}
\input{related}

\input{discussion}
\section*{Acknowledgements}
This work was supported by NSF grants CNS-1814105,
CNS-815219, and GRFP-1938052. 
Any opinions, findings, and conclusions or recommendations expressed in this material are those of the author(s) and do not necessarily reflect the views of the National Science Foundation.

We thank our reviewers and shepherd for their constructive feedback.

\appendix

\input{extraction_pseudocode}
\input{grammar}

\input{segmentation}
\input{fsmerrors}
\input{fsmfigures}
\input{synthesiserrors}

\bibliography{emnlp2020,crisn}
\bibliographystyle{IEEEtran}

\end{document}

%% file: abstract.tex

\begin{abstract}
Automated attack discovery techniques, such as attacker synthesis or model-based fuzzing, provide powerful ways to ensure network protocols operate correctly and securely. Such techniques, in general, require a formal representation of the protocol, often in the form of a finite state machine (FSM). Unfortunately, many protocols are only described in English prose, and implementing even a simple network protocol as an FSM is time-consuming and prone to subtle logical errors. Automatically extracting protocol FSMs from documentation can significantly contribute to increased use of these techniques and result in more robust and secure protocol implementations.

In this work we focus on attacker synthesis as a representative technique for protocol security, and on RFCs as a representative format for protocol prose description. Unlike other works that rely on rule-based approaches or use off-the-shelf NLP tools directly, we suggest a data-driven approach for extracting FSMs from RFC documents.
Specifically, we use a hybrid approach consisting of three key steps: (1) large-scale word-representation learning for technical language, (2) focused zero-shot learning for mapping protocol text to a protocol-independent information language, and (3) rule-based mapping from protocol-independent information to a specific protocol FSM.
We show the generalizability of our FSM extraction by using the RFCs for
six different protocols: BGPv4, DCCP, LTP, PPTP, SCTP and TCP. 
We demonstrate how automated extraction of an FSM 
from an RFC can be applied to the synthesis of attacks, with TCP and DCCP as case-studies. 
Our approach shows that it is possible to automate attacker synthesis
against protocols by using textual specifications such as RFCs.

\end{abstract}

%% file: intro.tex

\section{Introduction} 
Automated attack discovery techniques, 
	such as attacker synthesis or model-based fuzzing,
	provide powerful ways to ensure network protocols operate
	correctly and securely. 
Such techniques, in general, require a formal representation of the protocol,
	often in the form of a finite state machine (FSM).
Unfortunately, many protocols are only described in English prose,
	and implementing even a simple network protocol as an FSM
	is time-consuming and prone to subtle logical errors.
Automated attack discovery techniques are therefore infrequently
	employed in the real world 
	because of the significant effort required to
	implement a protocol FSM. Automatically extracting protocol FSMs from documentation can significantly contribute to increased use of these techniques and result in more robust and secure protocol implementations.

We observe that for network protocols there is an untapped  
	{\em resource of information} 
	available in the form of RFCs. 
With the recent interest in using data 
	to automatically solve problems in several fields,
	we ask the question: 
	``Can we leverage formal prose descriptions of protocols 
	  to improve protocol security?''

Given the inherent ambiguity of natural language text, extracting protocol information is not a straightforward task. The writers of protocol specifications often rely on human readers' understanding of context and intent, making it difficult to specify a set of rules  to extract information. This is by no means unique to the computer networks domain, and as a result, the natural language community  shifted its focus over the last decade to statistical methods that can help deal with such ambiguity. At the same time, one can not just apply ``off-the-shelf'' implementations of NLP tools combined in an ad-hoc way, as training such tools on poorly selected datasets will result in reduced performance and cause the resulting applications to be brittle.

Unlike other software, network protocols follow a specific pattern: they are described by messages and FSMs, they must meet temporal safety and liveness properties, and they follow a structured language.
Thus, NLP tools trained on such aspects of a protocol are likely to generalize on protocols with a similar structure. Unlike other NLP tasks where high precision is needed, protocol validation is more robust to noisy  NLP results because the ultimate result comes from protocol execution.

NLP techniques have been applied selectively in related problems. \textsc{WHYPER}~\cite{whyper} and \textsc{DASE}~\cite{dase} apply NLP techniques to
identify sentences that describe the need for a given permission in a mobile application description
and extract command-line input constraints from manual pages, respectively.
The work in~\cite{TM} used documentation and source code to create an ontology allowing the cross-linking of software artifacts represented in code and natural language on a semantic level.

Several other works looked at inferring protocol
specification -- based on network traces~\cite{comparetti2009prospex,wang2011inferring,caballero2009dispatcher,cho2010inference,cui2007discoverer},
using program analysis~\cite{kothari2008deriving,cho2011mace,lin2008automatic,caballero2007polyglot},
or through model checking~\cite{lie2001simple,corbett2000bandera}.
Comparetti \etal~\cite{comparetti2009prospex} infer
protocol state machines from observed network
traces by clustering messages based on the similarity of
message contents and their reaction to the execution.
Caballero \etal~\cite{caballero2009dispatcher}
extracts the protocol message format, given a trace of protocol messages. Cho \etal~\cite{cho2010inference} extracts the protocol
state machines from network traces with the help of a set of end-user
provided abstraction functions to generate an abstract alphabet out of trace
messages. The approach relies intensively on human expert input.

{\bf Our contribution.}
In this work we focus on attacker synthesis as a representative technique for protocol security, and on RFCs as a 
representative format for protocol description. Our goal is to close the automation gap between automated
protocol specification and security validation by extracting protocol FSMs from
the corresponding RFCs.  Unlike other works that rely on rule-based approaches or use off-the-shelf NLP tools directly, we suggest a {\em data-driven approach} for extracting information from RFC documents. Off-the-shelf NLP tools are typically trained over news documents, and when applied to technical documents that include many out-of-vocabulary words (i.e. technical terms), their performance degrades substantially. Rule-based systems, on the other hand, are developed to support information extraction based on the specific format of the textual input. Unfortunately, different RFC documents define variables, constraints, and temporal behaviors totally differently.  Moreover, RFCs follow no common document structure.  Machine-learning systems can deal with these challenges, however training such systems from scratch requires significant human effort annotating data with the relevant labels, which could be different for different protocols. We confront these challenges with a hybrid approach consisting of three key steps: (1) large-scale word-representation learning for technical language, (2) focused zero-shot learning for mapping protocol text to a protocol-independent information language, and (3) rule-based mapping from protocol-independent information to a specific protocol FSM. 

While RFCs are the main form for textual specification for protocols, they do not necessarily contain the complete specification, referred to as canonical FSM. There does not exist a one-to-one mapping between the textual specification and a canonical formal specification of the state machine, as canonical FSMs are created based not only on information contained in RFCs, but on input from experts with domain knowledge. This is a limitation for any statistical NLP approach. In this paper, we propose an alternative intermediary representation (i.e. a grammar) that can be used to recover partial state machines.

Our approach exploits the large number of technical documents found in online technical forums to train a deep learning model, capturing the properties of and interactions between technical terms. This process {\em does not require direct annotation}, and does not add to the human effort involved in building the model. Our zero-shot information extraction approach builds on that representation. Since each protocol consists of its own set of predicates and variables, we suggest a zero-shot approach in which we separate between protocols observed during training and testing. The model learns to identify and connect concepts relevant for the training protocols and at test time it is evaluated on extracting a set of symbols which were not observed at training. The output of that step creates an {\em intermediate representation} of conditions, operations and transitions, extracted from protocol text. The final step transpiles the intermediate representation into an FSM written in \textsc{Promela} code~\cite{holzmann1997model}.
We make the following contributions:
\begin{itemize}
\item We propose an embedding that allows us to learn network technical terms without the need to annotate data. To learn this embedding, we collected a set of 8,858 {\em unlabeled} RFCs from \texttt{ietf.org} and \texttt{rfc-editor.org} covering aspects of computer networking, including protocols, procedures, programs, concepts, meeting notes and opinions. These documents contain a total of 475M words.

\item We suggest and evaluate an NLP framework for the task of recovering FSMs from the RFCs, designed to adapt to previously unobserved protocols. We show the generalizability of our FSM extraction by using the RFCs for
six different protocols: BGPv4, DCCP, LTP, PPTP, SCTP, and TCP. As part of the NLP framework we propose a general-purpose abstraction for annotating the segments of text in RFC specification documents that describe the FSM for each of six network protocols. 
For example, of the 20 transitions in the TCP FSM, our NLP pipeline can extract 17, either correctly or partially so.

\item We demonstrate how automated extraction of an FSM
from an RFC can be applied to the synthesis of attacks, with TCP and DCCP as case-studies.
We find that even when the extracted FSM has errors, 
we can generate attacks that are confirmed on a canonical hand-written model of the same protocol. However, the quality of the extracted FSM impacts the accuracy of the attack synthesis.
For example, in the case of TCP, we can find attacks against only one property using our NLP pipeline, as opposed to against all four when using the canonical FSM.
\end{itemize}

The code is available at https://github.com/RFCNLP.

The rest of the paper is organized as follows. We discuss
attack synthesis
and NLP techniques in Section \ref{sec:bk}.  
Our grammar is described in Section \ref{sec:annotation},
	technical language embedding in Section \ref{sec:embedding}, 
	parsing in Section \ref{sec:parser}, 
	and FSM extraction in Section \ref{sec:fsmextraction}.
We present the TCP and DCCP attack synthesis case studies in Section \ref{sec:task}. We evaluate NLP components, FSM extraction, and
automated attack synthesis in Section \ref{sec:evaluation}. 
We present related work in Section \ref{sec:rel}. We discuss limitations and improvements in Section \ref{sec:disc}.

%% file: bk.tex

\section{Motivation and Our Approach}
\label{sec:bk}

In this section, we summarize the motivation of our work, the main challenges related to the extraction of FSMs from specification documents, and the way our approach is designed to circumvent these challenges. 

\subsection{Need for Automated FSM Extraction}
	
Automated attack discovery methods typically model the system under attack abstractly ---
	either implicitly, e.g. with a statistical representation, or 
	explicitly, e.g. with a finite state machine~\cite{jero2015leveraging,kang2019automated,korg}.
An FSM represents a program as a graph, where the nodes are program \emph{states}
	and the edges are \emph{transitions} (i.e. changes in state).
Recent work in the theory of security use FSMs to define what it means to attack~\cite{dullien2017weird,korg}.
Conversely, various attack discovery methods leverage FSMs 
	to compute attacks~\cite{jero2015leveraging,tcpwn_ndss_2018,chen2019devils,abbr_raid_2020,korg}.

Current attack finding~\cite{jero2015leveraging,tcpwn_ndss_2018,abbr_raid_2020} and attack synthesis techniques~\cite{korg} rely on a manually defined FSM specified by an expert. Anecdotally, there are reports where such FSMs were derived from code directly because specifications lacked such a description~\cite{abbr_raid_2020}.

\subsection{Challenges in FSM Extraction from RFCs}

One common way of specifying protocols is with RFCs.
 While RFCs provide some structure  that can be exploited for automated information extraction, it is not a straightforward task. An RFC describes in natural language, which is inherently ambiguous, the protocol's variables, states, and conditions for state transitions. Even for humans, creating a formal model from the text requires considerable domain expertise. From an NLP perspective, this is a specialized information extraction task, called \textit{semantic parsing}~\cite{10.1007/978-3-540-70939-8_28}, mapping the protocol text into structured information: the FSM. The mapping consists of multiple inter-dependent predictions, each extracting individual elements from the document, which together should capture the conditions and transitions of the FSM. Unlike traditional semantic parsing domains that operate over short texts, such as mapping a request in natural language to a command for a personal assistant (e.g., \textit{``set timer to 30 minutes''}), extracting an FSM requires processing multiple interconnected sentences to capture the transitions from just a single state.

Recently, very promising results were obtained by NLP researchers using deep learning methods for information extraction and semantic parsing tasks~\cite{cheng-etal-2017-learning,gardner-etal-2018-neural,chiu-nichols-2016-named,wadden-etal-2019-entity}.  However, most of the recent successes in these areas depend on large amounts of annotated data. When dealing with technical domains, high-quality annotated resources are scarce, and the fact that a deep understanding of the protocols is needed to annotate these documents makes generating enough data to support machine learning systems a costly and difficult process. 

Furthermore, dealing with specialized domains also reduces the amount of non-labeled data which can potentially be used. Unlike traditional NLP domains, such as newswire text, in which a vast amount of data is available for training NLP models, when learning to extract FSMs from network protocol RFCs, we are limited by the number of existing protocols. Given the data scarcity problem, it is difficult to build an NLP model that will reliably generalize to new protocols that were not observed during the training phase, as there is no common document structure to RFCs and the different functionality described by the protocols results in a different set of symbols and behaviors used by each protocol. In the next section we describe our approach for dealing with these challenges. 

\begin{figure}[ht]\centering
\includegraphics[width=0.5\textwidth]{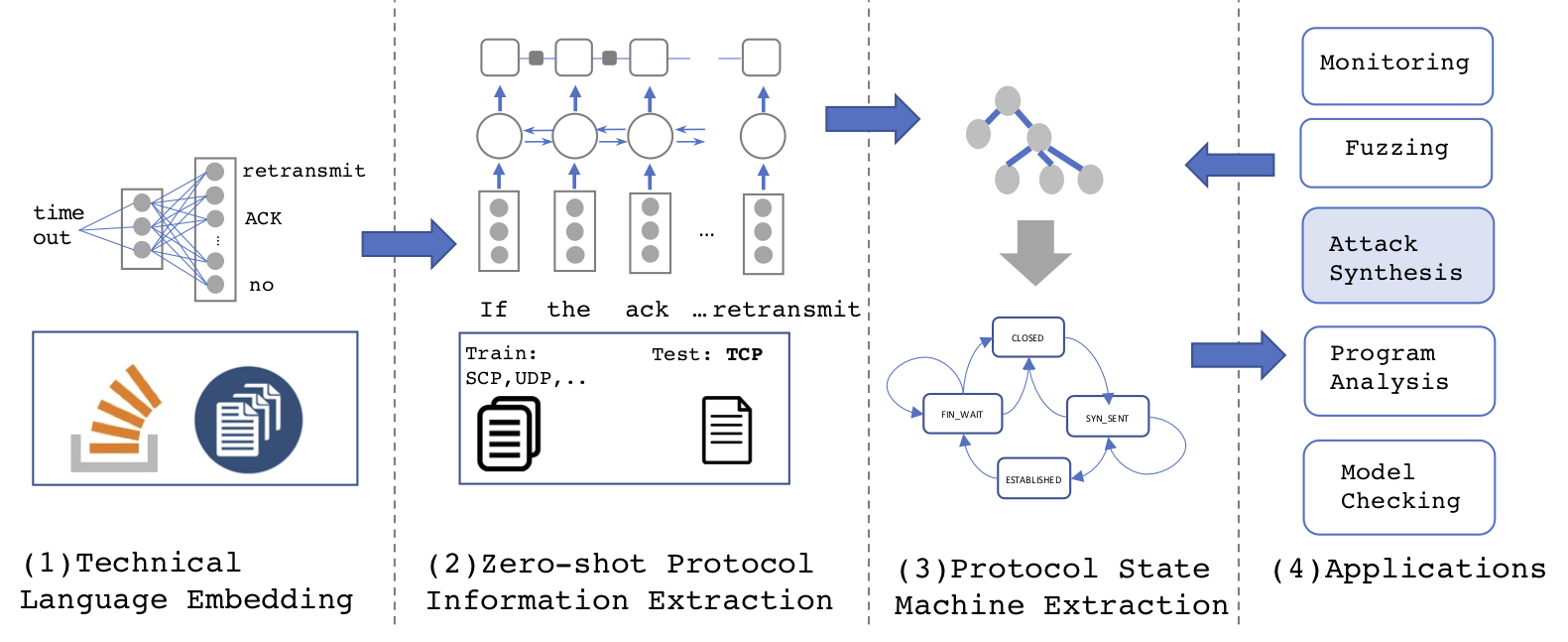}
\caption{Overview of our approach.}\label{fig:overview}
\end{figure}

\subsection{Our Approach}
As we discussed in the previous section, the lack of training resources prevents us from taking an end-to-end learning approach, in which a complex neural model is trained to predict the complete FSM directly from the text. Instead, we break the process into several parts,  allowing us to exploit several different forms of supervision and human expertise.  Figure \ref{fig:overview} describes our approach consisting of the following steps:

\textit{Technical Language Embedding.}
While there are only a few protocol RFCs, there are large amounts of technical documents discussing them and other related networking concepts that provide the background for understanding the RFC.  These  documents include technical forums, blogs, research papers, and specification documents. We exploit these documents to learn a distributed word representation, also known as an \textit{embedding} model, for technical language. The main advantage of this step is that it is an unsupervised process, and we do not require any annotations. Learning these representations will allow us to carry over information from the networking domain to our next step. Section~\ref{sec:embedding} describes this step in detail.

\textit{Zero-Shot Protocol Extraction.}
Once we have this representation, we turn our attention to learning a model to extract information regarding the FSM from the RFCs. To do this, we define a grammar that describes a higher-level abstraction of the structure of a general FSM for network protocols. While general, this abstraction will allow us to leverage different protocols to learn to extract this information, even when the underlying structure of their documents, the way the FSM is described and the specific names of variables, events and states vary between protocols. We explain this grammar in Section~\ref{sec:annotation}.   We annotate a set of six protocols, and use a zero-shot learning approach, in which the document for the predicted protocol is {\em completely unobserved during training}.  The output of this step
is a generic representation referred to as the {\em intermediate representation.} Section~\ref{sec:parser} describes this step in detail.   

\textit{Protocol State Machine Extraction.}
The extracted information structured according to our general protocol grammar must be converted into an actual FSM implementing the described protocol.  We use a set of heuristics to extract an FSM from the intermediate representation, as detailed in Section~\ref{sec:fsmextraction}.

%% file: annotation.tex

\section{Finite State Machine Grammar}\label{sec:annotation}

We define a general grammar to represent the state machine for the pertinent network protocols in their corresponding RFC specification documents. We use this grammar to annotate the segment of texts that describe the states, variables, and events that are relevant to the state machine, as well as the actions and the logical flow describing their behavior. Annotations are done using XML. We consider four
types of annotation tags: \emph{definition tags}, \emph{reference tags}, \emph{state machine tags}, and \emph{control flow tags}, which are formalized below. Finally, we formulate the grammar in Backus-Naur Form in Figure~\ref{fig:bnc_grammar}.

\subsection{Definition Tags}\label{sec:annotation_deftags}

Definition tags are used to annotate the names of states, events, and
variables that are relevant to each protocol. These are text segments
that are referenced throughout the document, and stake a role in
defining the state machine. For example, a message may be tagged as an
event if the receipt of such a message leads to a state transition. 

{\it State definition.} When the name of a state is introduced in the
text, it is annotated as such. Specifically, the \xml{<def_state>}
surrounds the first usage of the state name that is part of some
discernible pattern. Often this pattern is in the form of a newline-delimited list or bullet points, but can also appear as a comma-delimited list. The tag also includes an
identifier that is unique among the states. For example
\xml{<def_state id="##">IDLE</def_state>}, where \#\# is replaced by
the identifier. We assign state identifiers (SID) as monotonically
increasing integers in order of first appearance. Punctuation trailing
the state name is not included in the tag. These tags and SIDs will be
referenced by \emph{reference tags}. 

{\it Event definition.}  Events are also annotated to be referenced
throughout the text. Events follow the same annotation conventions as
states and use the annotation form: \xml{<def_event id="##">}. We will
refer to unique event identifiers as EIDs. 

{\it Variable definition.}  Variables are defined in a similar way to
events and states, however they do not include an analog to SIDs or
EIDs, because they are not explicitly referenced by annotation in the
rest of the text.

\subsection{Reference Tags}
When an event or state occurs in the text, it must be linked to an
event or state which was tagged. They need to be explicitly
defined because sometimes the proper name of a state or event will not
be used. For example, an RFC may formally refer to one event as
``ACK'', but throughout the text these ACKs may also be referred to as
``acknowledgments''. These are really the same event, and the
reference tags are used to clarify that. 

{\it State reference.} States are referenced by surrounding the
state's name throughout the text with the \xml{<ref_state id="##">}
tag, where \#\# corresponds to the appropriate SID that was included
with the state's \xml{<def_state>} tag. Punctuation trailing the state
name is not included in the tag. An example might look like the
following: \xml{enter <ref_state id="2">SYN-SENT</ref_state> state}. 

{\it Event reference.}  Events follow the same convention as state
references. The event reference must also include the \emph{type} of event, where the three possible types are: send, receive, and compute. Type tags are included as XML attributes, and will appear as in the following example: \xml{a <ref_event   type="send" id="10">SYN</ref_event> segment}.

\subsection{State Machine Tags}

We define a set of five tags to represent the state machine
logic. These are the crux of the annotation.

{\it Transition.} Denotes a state change that happens in the given context. We use argument tags \xml{<arg_source>}, \xml{<arg_target>} and \xml{<arg_intermediate>} to specify the segment in the text playing that role. For example, \xml{<transition>The server moves from the <arg_source>OPEN state</arg_source>, possibly through the <arg_inter>CLOSEREQ state</arg_inter>, to <arg_target>CLOSED</arg_target></transition>}. Note that in this case, the mentions to ``OPEN'', ``CLOSEREQ'' and ``CLOSED'' would also be enclosed in a \xml{<ref_state>} tag. In cases where the text is not explicitly annotated with argument tags, the states mentioned are assumed to be the ending states of the transition.

{\it Variable.} Certain variables may be tracked as part of the state machine. This
tag should be used to surround any logic that indicates that any of
these variables are altered or set to a new value. For example,
\xml{<variable>SND.UP <- SND.NXT-1</variable>} 

{\it Timer.} This tag is used if a timer value is changed or set. For example,
\xml{<timer>start the time-wait timer</timer>}. 

{\it Error.} If a context results in an error or warning being thrown, the error
message is then surrounded by this tag. For example,
\xml{<error>signal the user error: connection aborted due to user  timeout</error>}. 

{\it Action.} If a given context demands that some
action be taken, we use this tag. We specifically mark three types of actions: \textit{send}, \textit{receive} and \textit{issue}. Type tags are included as XML attributes. We use an argument tag \xml{<arg>} to specify the argument in the text being sent, received or computed. For example: \xml{<action type="send">Send <arg>a SYN segment</arg></action>}. Note that in this case, the mention to ``SYN" would also be enclosed in a  \xml{<ref_event>} tag: \xml{<ref_event id="10">SYN</ref_event>}. Additionally, there are certain events that are ambiguous in terms of how they
relate to the state machine, in those cases, this tag can be used without further specifications.

\subsection{Flow Control Tags}

A \xml{<control>} tag is introduced to indicate that some flow control
or conditional logic is about to follow. The flow control logic should
contain a \xml{<trigger>} tag, which captures the event that triggers
some action in the state machine, followed by one or more of the state
machine tags. A single block of control tags may contain multiple
state machine tags. These state machine tags should be in the form of
a list.  In this case, the implication is that the state machine tags
should all be executed if the initial trigger condition is true. Figure~\ref{fig:ann_example}
in Appendix \ref{app:grammarexample}
shows an example of a list of events within one control block from the TCP RFC (a.k.a. RFC 793)~\cite{rfc793}.

\subsection{Grammar}

Let \texttt{engl} denote any valid string in the English
language. Then, the grammar for the state machine annotation can be
described in Backus-Naur Form as observed in Figure
\ref{fig:bnc_grammar}. Here, \texttt{relevant=true} indicates that the
corresponding annotation is relevant to the protocol state-machine.

\begin{figure}[ht]	
\begin{lstlisting}
bool			::= true | false 
type			::= send | receive | issue
def-tag		::= def_state | def_var | def_event
ref-state ::= ref_state id="##"
ref-event ::= ref_event id="##" type="type"
ref-tag 	::= ref-event | ref-state
def-atom 	::= <def-tag>engl</def-tag>
sm-atom   ::= <ref-tag>engl</ref-tag> | engl
sm-tag 		::= trigger | variable | error | timer 
act-atom  ::= <arg>sm-atom</arg> | sm-atom
act-struct::= act-struct | act-struct act-atom
trn-arg   ::= arg_source | arg_target | arg_inter
trn-atom  ::= <trn-arg>sm-atom<trn-arg> | sm-atom
trn-struct::= trn-struct | trn-struct trn-atom
ctl-atom 	::= <sm-tag>sm-atom</sm-tag> 
           | <action type="type">act-struct</action> 
           | <transition>trn-struct</transition>
           | sm-atom
ctl-struct::= ctl-atom | ctl-struct ctl-atom
ctl-rel 	::= relevant=bool
control		::= <control ctl-rel>ctl-struct</control>
e 				::= control | ctl-atom | def-atom 
					 |  e_0 e_1
\end{lstlisting}
\caption{BNF grammar for RFC annotation.}
\label{fig:bnc_grammar}
\end{figure}

%% file: embedding.tex

\section{Technical Language Embedding}\label{sec:embedding}

In this section we describe our approach to learn distributed word representations for technical language. 
We start by providing some background about the techniques used to learn these representations, then we describe our embedding in detail.

\subsection{Background}

Distributed representations of words aim to capture meaning in a numerical vector. Unlike symbolic representations of words, that use binary values to signal if the words are present or not, word embeddings have the ability to generalize by pushing semantically similar words closer to each other in the embedding space. When using binary representations of words, we can only consider features that we have seen during training. Consider a scenario in which during training, we only have access to DCCP. If we were to test our learned model on TCP, we could not represent words that were not observed during training.

Several models have been suggested to learn distributed word representations. Some approaches rely on matrix factorization of a general word co-occurrence matrix~\cite{pennington2014glove}, while other approaches use neural networks trained to predict the context surrounding a word, and in the process, learn efficient word embedding representations in their inner layers~\cite{Mikolov13nips,peters-etal-2018-deep}. In this paper we focus on contextualized word representations. Unlike static word representations that learn a single vector for each word form, contextualized  representations allow the same word form to take different meanings in different contexts. For example, in the sentence ``The connection is in error and should be reset with Reset Code 5'', the word ``reset'' has two different meanings. Contextualized representations compute different vectors for each mention.

State-of-the-art pre-trained language models provide a way to derive contextualized representations of text, while allowing practitioners to fine-tune these representations for any given classification task.  One example of such models is BERT (Bidirectional Encoder Representations from Transformers) \cite{devlin-etal-2019-bert}. BERT is built using a Transformer, a neural architecture that learns contextual relations between words in a word sequence. A Transformer network includes two mechanisms, an encoder that reads the input sequence, and a decoder that predicts an output sequence. Unlike directional models that read the input sequentially, Transformer encoders read the whole sequence at once, and allow the representation of a given word to be informed by \textit{all} of its surroundings, left and right. Details regarding the Transformer architecture can be found in the original paper \cite{vaswani2017attention}.

\begin{figure}[h]
\includegraphics[width=\columnwidth]{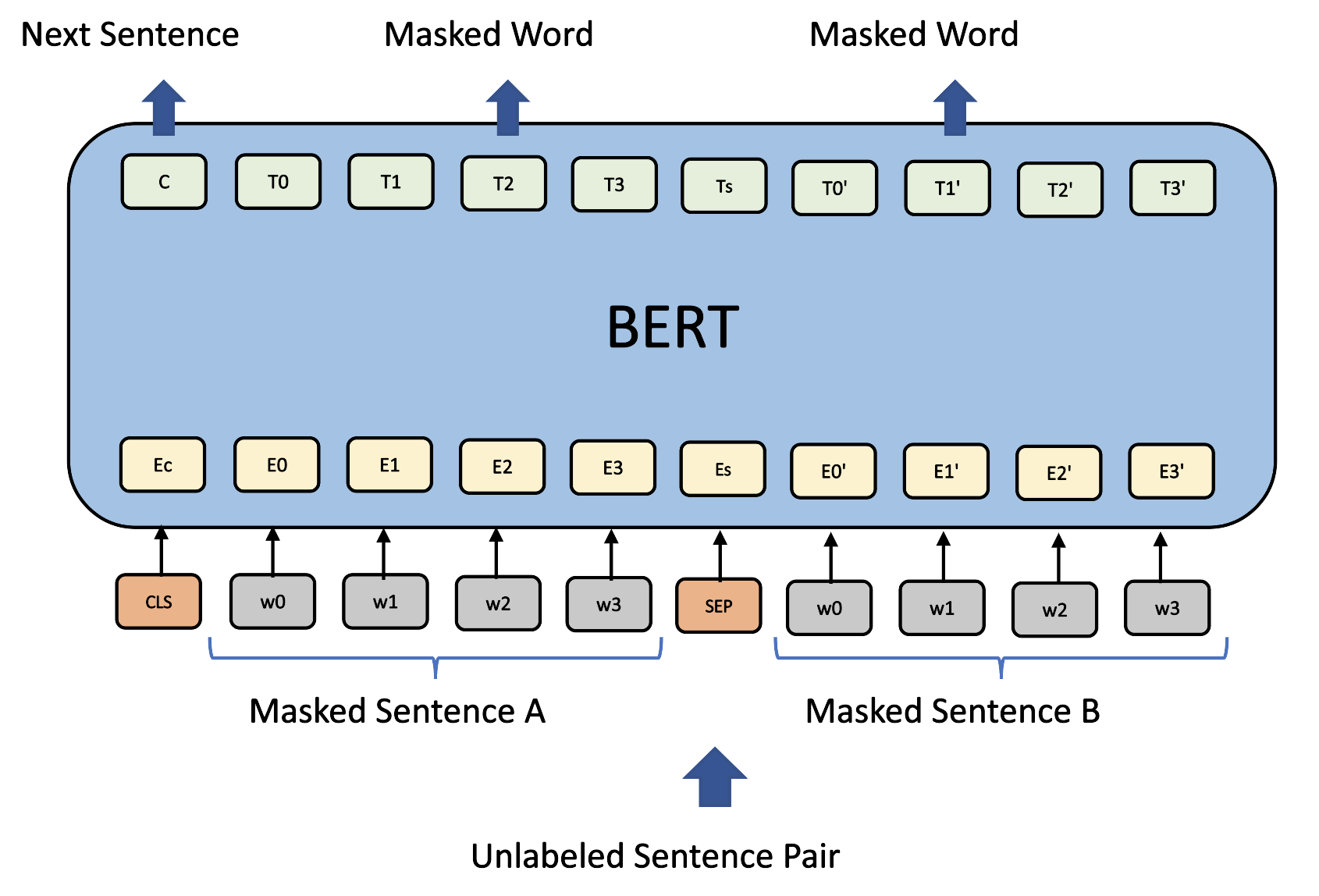}
\caption{BERT pre-training.}\label{fig:bert}
\end{figure}

To learn representations, BERT uses two learning strategies, masked language modeling  and next sentence prediction. The first strategy masks 15\% of the words in each sentence, and attempts to predict them. The second strategy uses pairs of sentences as input, and learns to predict whether the second sentence is the subsequent sentence in the original document.  Figure \ref{fig:bert} illustrates this process. BERT models were pre-trained on the BooksCorpus (800M words) and English Wikipedia (2,500M words) and are publicly available\footnote{https://github.com/google-research/bert}.

\subsection{Our Embedding}

While we could use pre-trained language models directly for predicting FSM tags, we note that these models were trained on general document repositories. To obtain a model that better represents the domain vocabulary, we further pre-train BERT using the masked language model and the next sentence prediction objective using networking data. We collected the full set of RFC documents publicly available in \texttt{ietf.org} and \texttt{rfc-editor.org}. These documents cover different aspects of computer networking, including protocols, procedures, programs, concepts, meeting notes and opinions. The resulting dataset consists of 8,858 documents and approximately 475M words. Note that we do not need any supervision for this  step.

Previous findings suggest that further pre-training large language models on the domain of the target task consistently improves performance \cite{gururangan-etal-2020-dont}. Our experiments in Section \ref{sec:evaluation} support this hypothesis.

%% file: parser.tex

\section{Zero-shot Protocol Information Extraction}\label{sec:parser}

In this section we describe our design for a protocol information extraction system. Our main goal is to have a system that can adapt to new, unobserved protocols without re-training the system. To support this, we build on the general grammar introduced in Section~\ref{sec:annotation} that focuses on concepts relevant to a wider set of protocols and takes advantage of the technical language embedding described in Section~\ref{sec:embedding}.

\subsection{Sequence-to-Sequence Model}

\begin{figure}[H]
    \centering
   \includegraphics[width=8.5cm]{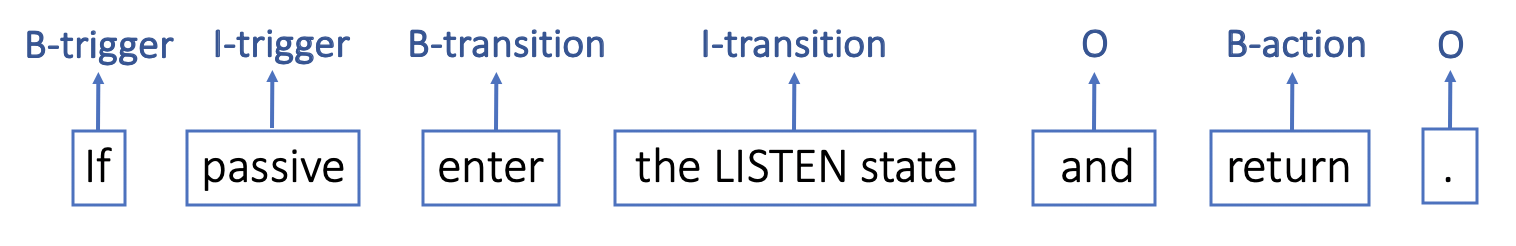}
    \caption{BIO example.}
    \label{fig:bio_example}
\end{figure}

To parse specification documents, we designed a sequence-to-sequence model that receives text blocks as input, and outputs a sequence of tags corresponding to the grammar described in Section~\ref{sec:annotation}. To tag the text, we use BIO (Beginning, Inside, Outside) tag labels. Text blocks correspond to paragraphs in the RFC document. Initially, we segment paragraphs into smaller units (e.g. individual words, chunks or phrases). Then, we map each unit to a particular tag. To illustrate this process, consider the parsed statement in Figure~\ref{fig:bio_example}, mapping chunks to BIO-tags.

We consider two models to learn the sequence to sequence mapping: 
a linear model we refer to as \linear, and a neural
model based on the BERT embedding, which we refer to as \bert.

Linear-Chain Conditional Random Fields (\linear) works on a set of extracted features over each chunk.  Conditional Random Fields model the prediction as a probabilistic graphical model; Chain Conditional Random Fields specifically consider sequential dependencies in the predictions~\cite{lafferty_crf}. 

Let $\bm{y}$ be a tag sequence and $\bm{x}$ an input sequence of textual units. We want to maximize the conditional probability:

\begin{equation}
\begin{split}
p(\bm{y}|\bm{x}) & = \frac{p(\bm{y}, \bm{x})}{\sum_{\bm{y'}} p(\bm{y'},\bm{x})} \\  
p(\bm{x}, \bm{y}) & = \prod^{T}_{t=1} \text{exp}(f (y_t, y_{t-1}, \bm{x_t}; \bm{\theta})) 
\end{split}
\label{eq:crf}
\end{equation}

\noindent Where $f$ is a linear scoring function learned with parameter vector $\bm{\theta}$ over a feature vector $\bm{x_t}$. To learn $\bm{\theta}$, we minimize the negative log-likelihood $-\log{p(\bm{y},\bm{x})}$. Learning is made tractable by using the forward-backward algorithm to calculate the partition function $Z(\bm{x}) = \sum_{\bm{y'}} p(\bm{y'},\bm{x})$.

\begin{figure}
   \centering
   \includegraphics[width=\columnwidth]{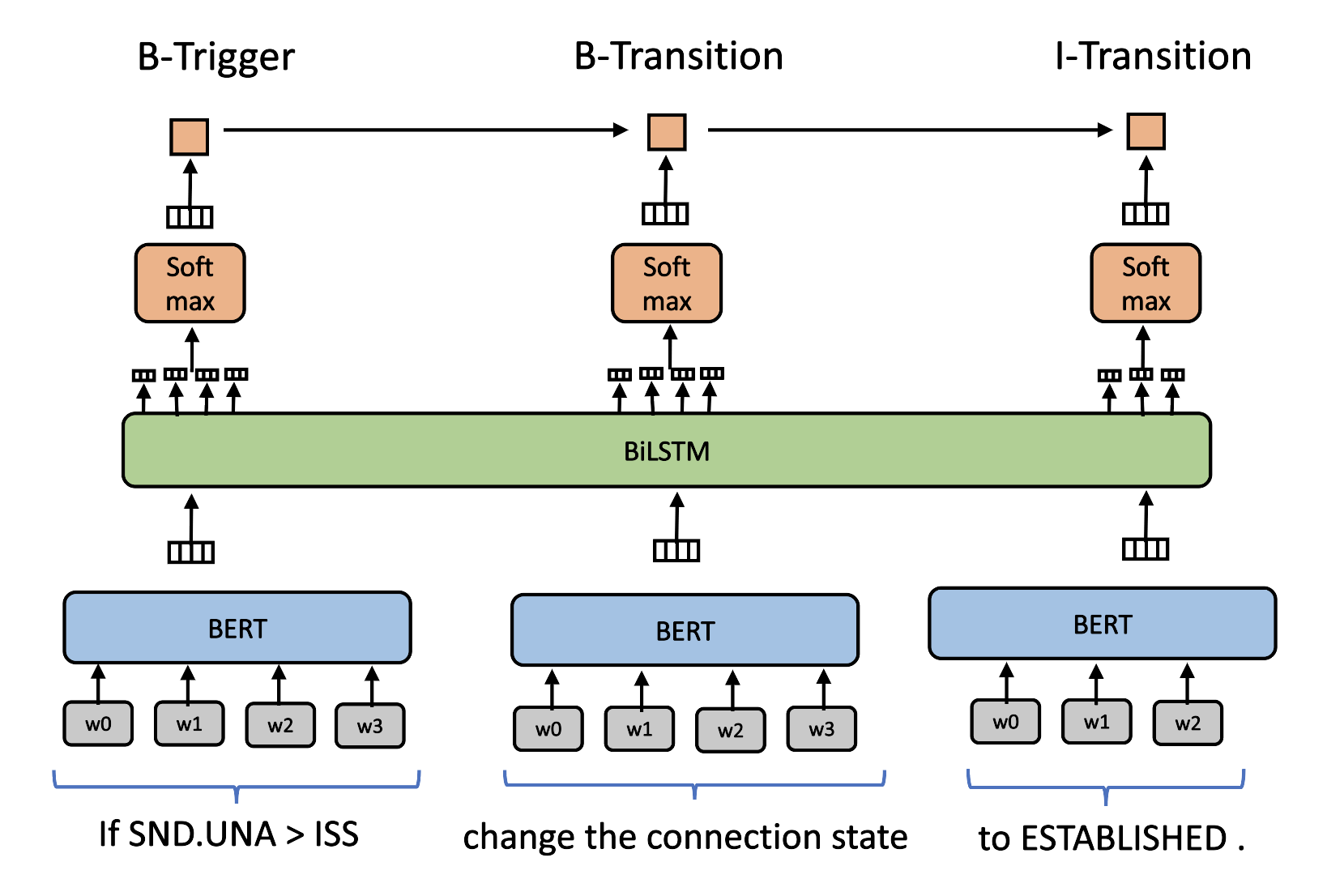}
    \caption{\bert. }
    \label{fig:bilstm_crf}
\end{figure}

The second model considered is a BERT encoder enhanced with a Bidirectional LSTM CRF layer (\bert). LSTMs are recurrent neural networks, a class of neural network where connections between nodes form a directed graph along a sequence~\cite{hochreiter1997long}. We outline this model in Figure~\ref{fig:bilstm_crf}. The BERT encoder is used to create chunk-level representations from word sequences. The resulting sequence of chunks is then processed using a BiLSTM. A softmax activation is used to obtain scores for the labels. Finally, we add a CRF layer on top. This way, we are able to leverage the sequential dependencies both in the representation and in the output space~\cite{lample-etal-2016-neural,ma-hovy-2016-end}. Note that BERT enforces a limit of 512 tokens per sequence, which is not enough to represent some of our control sequences. For this reason, we leverage a BiLSTM instead of using the CRF layer directly over BERT.

To formalize the {\bert} model, we first consider a textual unit containing $n$ words $(\bm{w_0, w_1, ..., w_{n-1}})$. A BERT encoder is used to obtain a single representation $\bm{u}$ for the full textual unit, resulting in a $d$-dimensional vector.

Then, a BiLSTM computes a representation over the sequence of embedded textual units $(\bm{u}_0, \bm{u}_1, ..., \bm{u}_{m-1})$ to obtain representations $\bm{h_t} = [\overrightarrow{\bm{h}_t}; \overleftarrow{\bm{h}_t}]$ for every textual unit $t$. Here, $\overrightarrow{\bm{h}_t}$ represents the left context of the sequence, and $\overleftarrow{\bm{h}_t}$ represents the right context, at every unit $t$.

Finally, we add a CRF layer over these representations by replacing the function $f$ in Eq.~\ref{eq:crf} with:

\begin{equation}
\begin{split}
f(y_t, y_{t-1}, \bm{x}_t) & = \bm{h_t} + \bm{P}_{y_t, y_{t-1}} 
\end{split}
\end{equation}

\noindent Where $\bm{x}_t$ represents the input for that textual unit, $\bm{h_t}$ is the representation of the textual unit computed with our model and $\bm{P}$ is a learned parameter matrix representing the transitions between labels. Like in the linear CRF case, we minimize the negative log likelihood, $-\log{p(\bm{y},\bm{x})}$, to jointly learn the parameters of the BERT encoder, the BiLSTM layer, and the transition matrix $\bm{P}$.

Predictions for both models are done using the Viterbi algorithm. Viterbi is a dynamic programming algorithm for finding the most likely sequence of states. Viterbi takes into account both emission ($\bm{h^2_t}$), and transition ($\bm{P}_{y_t,y_{t-1}}$) scores at each unit $t$ in the sequence.

\subsection{Features}\label{sec:features}

For each textual unit in the input, we extract a set of features to capture properties about the input and help us make a correct classification.

\textit{Vocabulary.}  We extract bag-of-word features for all stemmed forms of the words in the training data. Stemming is the process of producing morphological variants of a root word. This way we can reduce redundancy, as word stems and their inflected or derived words usually have the same meaning.
 
\textit{Capitalization Patterns.} We use features to indicate the different capitalization patterns of the original words (before stemming). We consider a feature for each of the following patterns: all letters are in lower case, all letters are capitalized, the first letter is capitalized, the word is in camel case, the word has only symbols, the word has only numeric characters, or the word has any other form of alpha-numeric capitalization.

\textit{Logical and Mathematical Expression Patterns.} We identify different patterns corresponding to logical and mathematical expressions. These include assignments ($x := a$, x $x \leftarrow a$, $x = a$), comparisons ($a < b$, $a > b$, $a \leq b$, $a \geq b$, $a == b$), and arithmetic and algebraic expressions.
 
\textit{Dictionary Features.} We include indicator features for a held-out dictionary of reserved state and variable names.

\textit{Part-of-Speech Tags.} We include part-of-speech (POS) tags for all observed words (e.g. noun, verb, adjective). For extracting POS tags, we use an off-the-shelf tagger.

\textit{Position Features.} We use position and relative position indicators for each word in a chunk.

All of the features used are standard in general NLP pipelines. For the \linear, this collection of features represents the input $\bm{x_}t$ for each textual unit $t$.  For the \bert, we concatenate the feature vector to the resulting vector $\bm{u}_t$ from the BERT encoder, before being inputted to the BiLSTM layer.

\subsection{Post-Processing}\label{sec:postproc}

We experiment with a set of \textit{rules} to correct some easy cases that the prediction models fail to identify. The rules are applied on top of the classification output, by flipping labels in the relevant cases. First, we look for textual units with mentions to states. If the unit mentions a state, and there is a transition verb (e.g. move, enter) or a directional preposition (e.g. to, from), we label the unit as a \textit{transition} span. Then, we look for textual units with mentions to events. If the unit mentions an event, and there is an action verb (e.g. send, receive), we label the unit as an \textit{action} span. Then, we label any remaining unlabeled span with mentions to states or events as a \textit{trigger}. Finally, we label any remaining unlabeled spans with mentions to variable names, ``error'' or ``timer'' as \textit{variable}, \textit{error} and \textit{timer}, respectively. We refer to the models that use these rules as \linearrules ~and \bertrules.

Once the triggers, transitions, actions, variables, and errors are identified, we use an off-the-shelf Semantic Role Labeler (SRL)~\cite{Gardner2017AllenNLP} to identify predicted actions as either send, receive, or issue, depending on the verb used, as well as to identify the segment in the text being sent, received, or issued. Semantic Role Labeling consists of detecting semantic arguments associated with the predicate or verb in a sentence, and their classification into their specific roles. For example, given a sentence like ``Send a SYN segment'', an SRL model would identify the verb ``to send'' as the predicate, and ``SYN segment'' as the argument. Identified arguments are then tagged using the \texttt{<arg>} tag introduced in Section~\ref{sec:annotation}. We also use the SRL output to identify transitions verbs such as enter and leave, and identify the segment in the text being explicitly mentioned as the source or target for this transition. For example, in the sentence ``client and server sockets enter this state from PARTOPEN'', the SRL model identifies the verb ``to enter'' as the predicate, the segment ``this state'' as the argument and ``from PARTOPEN'' as the directional modifier. Arguments and directional modifiers are then tagged as as \texttt{<arg\_source>}, \texttt{<arg\_target>} or \texttt{<arg\_intermediate>}, depending on the prepositions used.

In addition, we use exact lexical matching to identify explicit mentions to states and events in the predicted sequences. We keep track of the indentation in the original documents to infer the scope of \texttt{<control>} statements. The resulting tagged text constitutes the {\xmlrep} that will be used for extracting the \fsm.

%% file: translation.tex

\section{FSM Extraction}\label{sec:fsmextraction}

The \xmlrep~obtained using our \linear~or \bert~model is not an \fsm, thus we need a procedure to
extract an \fsm ~from the \xmlrep.
The {\fsm} is expressed as
\(P = \langle S, I, O, s_0, T \rangle\)
	with finite \emph{states}~$S$, 
		 finite \emph{inputs}~$I$, 
		 finite \emph{outputs}~$O$ disjoint from~$I$, 
		 \emph{initial state}~$s_0~\in~S$, and
		 finite \emph{transitions} 
		 $T \subseteq S \times ( \{ \epsilon, \textit{timeout} \} \cup (I \cup O)^* ) \times S$.

We extract the states $S$ by scanning the \xmlrep~for \texttt{def-state}s.
If one of the \texttt{def-state}s
 contains ``initial" or ``begin" in its body,
	we set the corresponding state as the initial state~$s_0$;
	otherwise we just choose whichever is the first \texttt{def-state} in the document.
We extract the inputs~$I$ and outputs~$O$ by scanning for \texttt{def-event}s
	where the \texttt{type} is \texttt{receive} or \texttt{send}, respectively.

Although the \xmlrep~contains \texttt{transition} blocks, these blocks do not exactly contain actual \fsm~transitions.  Rather, they contain pointers for  \emph{where to look} in the \xmlrep~in order to guess the source and target states, and labels, for the \fsm~transitions.  A \texttt{transition} block might describe no transitions at all, or multiple transitions at once.  It might describe only part of a transition, for example the label and the target state, while the rest is described somewhere else in its context.  Such cases can occur even with a perfect \xmlrep, because of complex syntax and formatting used in the RFC text. To obtain the transition set $T$ 
we proceed in two steps: first we extract potential transitions from the \texttt{transition} blocks; then we heuristically prune transitions that look like noise.

\emph{Potential Transition Extraction.}  We define an initially empty set of \emph{possible transitions} $T_{\textit{pos}}$.  For each \texttt{transition} block~\textsf{T} in the \xmlrep~\textsf{xml}, we compute potential transitions described in~$\textsf{T}$ using the Algorithm~\textsc{extractTran}.  Briefly, \textsc{extractTran} searches \emph{lower} in the \xmlrep~to find target states, and \emph{higher} to find source states.  It handles sentences like ``starting at any state other than \texttt{CLOSED}'' using the set complement.  It also handles explicitly labeled intermediate states, so that sentences like ``the machine goes to \texttt{CLOSED}, then \texttt{REQUEST}, then \texttt{PARTOPEN}'' are interpreted as $\texttt{CLOSED} \xrightarrow[]{} \texttt{REQUEST} \xrightarrow[]{} \texttt{PARTOPEN}$ rather than $\texttt{CLOSED} \xrightarrow[]{} \texttt{PARTOPEN}, \texttt{REQUEST} \xrightarrow[]{} \texttt{PARTOPEN}$.  It uses the helper function \textsc{extractTranLbl} to guess the transition label $\ell$, recursing upward in the ancestry of \textsf{T} at most six times until the result is well-formed.  
Pseudocode for \textsc{extractTran} is given in the Appendix.

\emph{Heuristic Transition Pruning.}  After adding the potential transitions extracted from each \texttt{transition} block \textsf{T} to a set $T_{\textit{pos}}$, we filter $T_{\textit{pos}}$ using three heuristics.  First, we remove any possible transition $t \in T_{\textit{pos}}$ that does not type-check, that is, for which $t \notin S \times ( \{ \epsilon, \textit{timeout} \} \cup (I \cup O)^* ) \times S$.  Second, we apply a ``call and response'' heuristic, where if $T_{\textit{pos}}$ contains some transitions $s \xrightarrow[]{x? y!} s'$, $s \xrightarrow[]{x?} s'$, and $s \xrightarrow[]{y!} s'$, then the latter two are discarded because they are likely noise generated by the first one.  Third, we apply a ``redundant epsilons'' heuristic, where if $T_{\textit{pos}}$ contains some transitions $s \xrightarrow[]{\epsilon} s'$ and $s \xrightarrow[]{\ell} s'$, where $\ell \neq \epsilon$, then the $\epsilon$-transition is discarded because it is likely noise generated by the $\ell$-transition. 
The transitions $T$ is the remaining filtered set $T_{\textit{pos}}$.

%% file: task.tex

\section{Task: Attacker Synthesis }
\label{sec:task}

In this section we use attacker synthesis as an exemplifying application
for FSM extraction.

\subsection{Attacker Synthesis}

\emph{LTL program synthesis}, 
	also known as the \emph{LTL implementability problem},
	is to deduce for an LTL property $\phi$
	if there exists some program $P$ that makes $\phi$ true.
For example, $\phi$ could be the homework assignment
	to implement multi-\textsc{Paxos},
	and the program synthesis problem would be to automatically
	compute a satisfying code submission.
The problem is known to be doubly exponential 
	in the size of the property~\cite{pnueli1989synthesis}.

\emph{LTL attacker synthesis} is slightly different.
In this work we consider a centralized attacker synthesis problem for protocols,
	where the attacker has just one component.
Other variations on the problem are formulated 
	in~\cite{korg}.
Suppose $P \parallel Q$ is a system consisting of some programs $P$ and $Q$, and $\phi$ is an LTL correctness property which is made true by the system; that is $P \parallel Q \models \phi$.
Consider the threat model where $Q$ is the vulnerable part of the system.  The attacker synthesis problem is to replace $Q$ with some new \emph{attacker} $A$ having the same inputs and outputs as $Q$, such that the augmented system behaves incorrectly, that is, $P \parallel A$ violates $\phi$.  We only consider attackers which succeed under the assumption that (a) the attack eventually terminates, and (b) when the attack terminates, the vulnerable program $Q$ is run.
The \emph{program synthesizer} must compute a program
		that satisfies~$\phi$
		in all of its (non-empty set of) executions,
	but the \emph{attacker synthesizer} 
		only needs to compute a program that violates~$\phi$
		in one execution.

\subsection{Attacker Synthesis with \textsc{Korg}}

\textsc{Korg} is an open-source attacker synthesis tool for protocols.
It requires three inputs:
	(1) a \textsc{Promela} program $P$ representing the invulnerable part of the system;
	(2) a \textsc{Promela} program $Q$ representing the vulnerable part of the system, as well as its interface (inputs and outputs) in YAML format; and
	(3) a \textsc{Promela} LTL property $\phi$ representing what it means for the system to behave correctly.
\textsc{Korg} computes $\exists$-attackers (attackers for which there exists a winning execution) by reducing the attacker synthesis problem to a model-checking problem over the system $P \parallel \textsc{Daisy}(Q)$, where the vulnerable program $Q$ is replaced with a nondeterministic search automaton (called a Daisy Gadget) 
having the same interface
as $Q$.
The model-checker then computes
an execution that violates the correctness property,
and \textsc{Korg} projects the component of the execution representing
the gadget's actions into a new \textsc{Promela} program, which is the
synthesized attacker~\cite{korg}.

\begin{figure}[h]
\begin{adjustbox}{max totalsize={.45\textwidth}{.99\textheight},center}
\begin{tikzpicture}
\fill [orange] (-3,-1) rectangle (1,1.5);
\fill [cyan] (-3,2.5) rectangle (1,1.5);

\node[] (model) at (-1,2) {\textsc{Promela} program $P$};
\node[text width=100] (placement) at (-0.8,0.8) {\textsc{Promela} vulnerable program~$Q$};
\node[text width=100] (phi) at (-0.8,-0.3) {\textsc{Promela} LTL correctness property~$\phi$};

\node[draw,rectangle,fill=white,double] (korg) at (3,1) {\textsc{Korg}};

\draw[->] (1,2) -- (korg);
\draw[->] (1,0.8) -- (korg);
\draw[->] (1,-0.3) -- (korg);

\node[draw,rectangle,fill=white,double] (spin) at (7,1) {\textsc{Spin}};

\draw[->] (korg) to[above,out=north east,in=north west] node {``$P \parallel \textsc{Daisy}(Q) \models \psi$?"} (spin);

\draw[->] (spin) to[below,out=south west,in=south east] node {Counterexamples} (korg);

\node[] (attacks) at (3,-1) {Synthesized Attackers};

\draw[->] (korg) -- (attacks);
\end{tikzpicture}
\end{adjustbox}
\caption{\textsc{Korg} work-flow.  With our NLP pipeline, the user need only supply the orange inputs and the system RFC.  The property $\psi$ is automatically computed from $\phi$ to ensure the attacker eventually terminates, at which point the original code $Q$ is run.  The \textsc{Daisy} gadget is defined in~\cite{korg}.}
\label{fig:korg}
\end{figure}
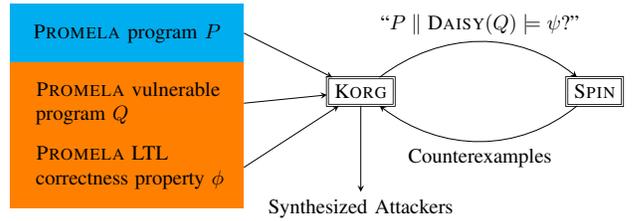

\subsection{TCP and DCCP Attacker Synthesis with \textsc{Korg}}

We focus on the TCP and DCCP connection establishment and tear-down routines
	as representative protocols for attacker synthesis.
The TCP connection routine was previously studied 
	using the attacker synthesis tool 
		\textsc{Korg} (Fig.~\ref{fig:korg});
	now we conduct a similar analysis for both TCP and DCCP using the same tool, 
	but we automatically
	extract FSMs using NLP.
We want to show that the FSMs extracted from our NLP pipeline
	can be used directly for attacker synthesis,
	alleviating the considerable engineering effort required to hand-model
	the system under attack.
We show the effectiveness of the FSM extraction on this task
in Section~\ref{sec:eval_synthesis}.

Our NLP pipeline and FSM extraction produce an FSM.
In order to use the extracted FSM for attacker synthesis, we transpile it to \textsc{Promela}.  For example, if we begin with the TCP RFC, then the result will be a \textsc{Promela} program describing the TCP connection routine.  

For each of TCP and DCCP, we hand-write four LTL properties in
	\textsc{Promela} based on a close reading of the corresponding RFC. Our TCP
	properties are given in Equation~\ref{eqn::tcp_props}, and our DCCP properties
	are given in Equation~\ref{eqn::dccp_props}.  We define the vulnerable
	\textsc{Promela} program $Q$ to be a generic message channel between peers.
	For each of the four $\phi_i$, we feed the inputs
	$P, Q,$ and $\phi_i$ to \textsc{Korg} and generate attackers. But how do we know if these attackers are legitimate, since they were generated with a potentially incorrect program $P$?  We solve this by testing the attackers against a Canonical \textsc{Promela} program. For TCP
we adapt the Canonical program from~\cite{korg}.  For DCCP, no such program was available and we created our own hand-written Canonical \textsc{Promela} program. 
\vspace{-10pt}

\begin{equation}
\begin{aligned}
\phi_1 = &\,\parbox[t]{20em}{``No half-open connections."} \\
\phi_2 = &\,\parbox[t]{20em}{``Passive/active establishment eventually succeeds."} \\
\phi_3 = &\,\parbox[t]{20em}{``Peers don't get stuck."} \\
\phi_4 = &\,\parbox[t]{20em}{``\texttt{SYN\_RECEIVED} is eventually followed by \texttt{ESTABLISHED}, \texttt{FIN\_WAIT\_1}, or \texttt{CLOSED}."}
\end{aligned}
\label{eqn::tcp_props}
\end{equation}

\begin{equation}
\begin{aligned}
\theta_1 = &\,\parbox[t]{20em}{``The peers don't both loop into being stuck or infinitely looping."} \\
\theta_2 = &\,\parbox[t]{20em}{``The peers are never both in \texttt{TIME\_WAIT}."} \\
\theta_3 = &\,\parbox[t]{20em}{``The first peer doesn't loop into being stuck or infinitely looping."} \\
\theta_4 = &\,\parbox[t]{20em}{``The peers are never both in \texttt{CLOSE\_REQ}."}
\end{aligned}
\label{eqn::dccp_props}
\end{equation}

Note that \textsc{Korg} expects that all its inputs ($P$, $Q$, and~$\phi$) are correct.
However, since we test on an automatically extracted \textsc{Promela} program $P$, 
	which may have some incorrect transitions when compared to the corresponding Canonical program,
	this assumption is broken.
We therefore adapted \textsc{Korg} to work on incomplete or imperfect programs, while preserving the 
formal guarantees from the original paper (except for soundness, which depends on how different the extracted program is from the Canonical one).

%% file: evaluation.tex

\section{Evaluation}\label{sec:evaluation}
In this section we present an evaluation of
both NLP tasks and attacker synthesis. 

We use ``Gold intermediary representation'' to refer to the manual text annotations obtained using our protocol grammar presented in Section III.
We use ``Canonical FSM'' to refer to the FSM which was derived from expert domain knowledge, the protocol RFC, and FSM diagrams in textbooks and literature.

\subsection{Information Extraction Evaluation}
 We evaluate how much of the \xmlrep~specified in Section~\ref{sec:annotation} we can recover.
\subsubsection{Methodology}
We evaluate the output of the specification document parser in six different protocols: BGPv4, DCCP, LTP, PPTP, SCTP and TCP.  We use a leave-one-out setup, by training on five protocols and testing on the remaining one. This means that no portions of a test protocol are observed during training. To artificially introduce more training sequences, we split recursive control statements into multiple statements at training time. At test time, we evaluate on each example once. 

We evaluate predictions at the token-level and at the span-level. For tokens, we have 19 labels: beginning and inside tags for trigger, action, error, timer, transition and variable, as well as the outside label. We use standard classification metrics to measure the token-level prediction performance. We infer the control spans based on the indentation in the original documents. For identifying event and state references, we do direct lexical matching using a dictionary built on the definition tags described in Section~\ref{sec:annotation_deftags}.

To evaluate spans, we use the metrics introduced for the International Workshop on Semantic Evaluation (SemEval) 2013 task on named entity extraction~\cite{segura-bedmar-etal-2013-semeval}. We use the SemEval evaluation script on our data. In this case, we have six span types, plus the outside tag. The metrics are outlined below.

\begin{enumerate}
\item \textbf{Strict} matching, with exact boundary and type.
\item \textbf{Exact} boundary matching, regardless of the type.
\item \textbf{Partial} boundary matching, regardless of the type.
\item \textbf{Type} overlap between the tagged span and the Gold span.
\end{enumerate}

We use the \linear~provided by the pystruct library \cite{JMLR:v15:mueller14a}, which uses a structured SVM solver using Block-coordinate Frank-Wolfe \cite{RePEc:wly:navlog:v:3:y:1956:i:1-2:p:95-110}, and use the default parameters during training. We implemented the \bert~model using the transformers library~\cite{wolf-etal-2020-transformers} and PyTorch \cite{pytorch-citation}, and learn the model using the adaptive gradient algorithm Adam, with decoupled weight decay \cite{loshchilov2018decoupled}. We initialize the BERT encoder with the parameters resulting from our pre-training stage described in Section \ref{sec:embedding}, which further pre-trains BERT on technical documents. We use a learning rate of 2e-5 and 50 hidden units for the BiLSTM layer. For BERT, we use the standard parameter settings, and a maximum sequence length of 512. We randomly sample 10 percent of the training data to set aside as a development set, which we use to perform early stopping during training, using a patience of 5 epochs.

\begin{table}
	\caption{Average Results for Different Models}\label{tab:different_models}
	\resizebox{\columnwidth}{!}{%
	\begin{tabular}{| l | c | c |  c | c | c | }
		\hline
		\multirow{2}{*}{Model} & \multicolumn{3}{|c|}{Token-level } & \multicolumn{2}{c|}{Span-level } \\ \cline{2-6}
~ &		Acc & Weighted F1 & Macro F1 &  Strict & Exact \\ \hline
Rule-based & 31.08
								   &  25.94 
								   & 29.37 
								   & 41.58 
								   & 41.78 \\
								   
BERT-base 
				& 58.93
				& 56.72
				& 51.33
				& 60.77
				& 84.18 \\
				
BERT-technical 
				& 62.38 
				& 60.31
				& 52.50
				&	62.84
				&	83.81 \\
				
\linear  & 58.95
				 & 56.61 
				 & 49.58 
				 & 63.98
				 & 85.65  \\
				 	 
\linearrules & 58.60
								   &  56.79 
								   & 50.62 
								   & 63.52 
								   & 85.18 \\

\bert 
				& \textbf{64.42}
				&	\textbf{64.18} 
				&	\textbf{54.95}
				&	\textbf{68.81}
				& \textbf{86.83} \\
				
\bertrules & 62.79
								   &  62.50 
								   & 53.64 
								   & 66.22 
								   & 86.10 \\			 

\hline
	\end{tabular}}
\end{table}

\subsubsection{Segmentation strategies}
We evaluate different segmentation strategies to create the base textual unit in our sequence-to-sequence models: segmenting by token, chunk, and phrase. For segmenting chunks, we use an off-the-shelf chunker~\cite{openNLP}. For segmenting phrases, we split the text on periods, colons, semi-colons and newline markers, as well as on a set of reserved words corresponding to conditional statements (e.g. if, then, when, while). We find that segmenting by chunks yields the best token-level results (Weighted F1 of 61.25), but segmenting by phrases gives us better span-level results (Strict matching of 63.98, and Exact boundary matching of 85.56). Detailed results for these models are in the Appendix, in Table~\ref{tab:segmentation_strategies}. Moving on, all evaluations are done using the phrase segmentation strategy. 

\subsubsection{Extraction models}

We evaluate the two models proposed in Section~\ref{sec:parser}, and obtain a significant improvement with respect to a rule-based baseline that applies the rules outlined in Section~\ref{sec:postproc} directly, without any learning. In addition, we test a BERT model by removing the BiLSTM CRF Layer, both with and without the pre-training strategy introduced in Section~\ref{sec:embedding}. Average results can be observed in Table~\ref{tab:different_models}. When pre-training on technical documentation is not done, we use the BERT model trained on BookCorpus and Wikipedia. Here, we can appreciate both the advantage of the technical embeddings, as well as the advantage of the BiLSTM CRF layer.
We find that leveraging expressive neural representations for sequence-to-sequence models is advantageous for this task. Note that both the \bert~and the \linear~models make use of the full set of features introduced in Section \ref{sec:features}. Finally, we find that applying rules on top of our models to correct predictions does not improve their general performance.

In Table~\ref{tab:crf_models}, we show the individual performance for the six protocols 
and show that we obtain better performance using the \bert~model for all protocols.

\begin{table}[ht]
\centering
	\caption{Results by Protocol for our Best Models}\label{tab:crf_models}
	
	\begin{tabular}{| l | c  c |  c  c | c |}
		\hline
		\multirow{2}{*}{Protocol} & \multicolumn{2}{|c|}{\linear} & \multicolumn{2}{c|}{\bert } & \# Control\\ 
		\cline{2-5}
~ & Strict & Exact & Strict & Exact & Statements  \\		 \hline
BGPv4 
	& 52.99 
	&	82.56 
	& \textbf{57.34}
	& \textbf{86.86}
	& 6\\
DCCP 
	& 69.74 
	&	92.73 
	& \textbf{75.60}
	& \textbf{93.25}
	& 150\\
LTP 
	& 67.25 
	&	\textbf{94.44}
	& \textbf{74.22}
	& 94.41 
	& 65 \\
PPTP 
	& 84.21 
	& 96.05 
	& \textbf{87.34} 
	& \textbf{98.73} 
	& 25 \\
SCTP 
	& 52.21 
	& 65.49 
	& \textbf{58.54} 
	&	\textbf{65.85} 
	& 19 \\
TCP 
	& 57.46 
	& \textbf{82.64} 
	& \textbf{59.82}
	&	81.90
	& 31 \\
\hline
	\end{tabular}
\end{table}

\subsubsection{FSM extraction}
We compare both the NLP and the Gold extracted FSMs with the Canonical FSM in Table \ref{tab:transition_extraction_comm}, based on how many states and transitions are successfully extracted. Both the NLP and the Gold FSMs are extracted from the predicted/annotated intermediary representation introduced in Section \ref{sec:annotation}, by using the procedure outlined in Section \ref{sec:fsmextraction}. The results presented in Tables I and II correspond to how accurately we can recover this intermediary representation from the text, before we attempt to construct the FSM.

Note that even with Gold annotations, we are not able to extract all expected transitions because in some cases, the transitions are not explicit in the text or in other cases, our general grammar and extraction procedure are not able to capture the intended behavior. In all cases, we are able to recover all relevant states. Graphic visualizations for all resulting state machines can be found in the Appendix.

We manually analyzed all the partially correct, incorrect and missed transitions in Table III
and found that, for the Gold FSM,  they are caused  by ambiguities in the RFC, or the information about some transition missing completely (67\% for TCP and 96\% for DCCP). The remaining errors are due to difficulties capturing complex logical flows using our method. The difference between the Gold FSMs and the predicted FSMs can always be attributed to errors in the text predictions.

\begin{table}
\caption{Transitions Extracted (Partially Correct means source and target state are correct, and at least one of the events on the edge is also correct).}
\label{tab:transition_extraction_comm}
\resizebox{\columnwidth}{!}{%
\begin{tabular}{|l|c|c|c|c|c|c|}
\hline
\multirow	{2}{*}{TCP FSM} & \multirow	{2}{*}{Canonical} & \multirow	{2}{*}{Extracted} & \multirow	{2}{*}{Correct} & Partially & \multirow	{2}{*}{Incorrect} & \multirow	{2}{*}{Not Found}  \\
~ & ~ & ~ & ~ & Correct & & \\
\hline
Gold
	&  \multirow	{5}{*}{20} 
	  & 18
    & 8
    & 8
    & 2 
    & 4 \\
    
\linear 
	& ~
	& 28 
	& 2  
	& 3 
	& 23 
	& 15 \\
    
\linearrules
	& ~
	& 30 
	& 7 
	& 10
	& 13 
	& 3 \\
	
\bert 
	& ~
	& 11 
	& 2 
	& 3
	& 6 
	& 15 \\
	
\bertrules
	& ~
	& 30
	& 7 
	& 10
	& 13 
	& 3 \\
\hline
\multirow	{2}{*}{DCCP FSM} & \multirow	{2}{*}{Canonical}  & \multirow	{2}{*}{Extracted} & \multirow	{2}{*}{Correct} & Partially  &\multirow	{2}{*}{Incorrect} & \multirow	{2}{*}{Not Found}  \\
~ & ~ & ~ & ~ & Correct & & \\
\hline
Gold
	& \multirow{5}{*}{34} 
	& 24
	& 15
	& 1
	& 8 
	& 18 \\
	
\linear 
	& ~
	& 8 
	& 1 
	& 5
	& 2 
	& 28 \\

\linearrules
	& ~
	& 17
	& 6 
	& 3 
	& 8 
	& 25 \\
	
\bert
	& ~
	& 20
	& 9 
	& 1 
	& 10
	& 24 \\	
	
\bertrules
	& ~
	& 19 
	& 8 
	& 3
	& 8
	& 23 \\
\hline

\end{tabular}}
\end{table}

For example, we notice that one incorrect behavior in the TCP Gold
  FSM is caused by ambiguity in the TCP RFC text.
The only outgoing communication transition in the TCP Gold FSM from \texttt{SYN\_SENT} 
	sends \texttt{ACK}
	and goes to \texttt{SYN\_RECEIVED}. 
The correct logic is to \emph{receive} \texttt{SYN} first,
	before sending the \texttt{ACK} and transitioning.
The TCP RFC 
	does not textually mention the expected \texttt{SYN}.
We only know to expect it because it is illustrated 
	in Figure 6 of the RFC.
We show more examples of FSM extraction errors in the Appendix.

\subsubsection{Summary} 

In Tables I and II we evaluated how much of our intermediary representation we could extract from natural language, while in Table III we evaluated how much of the canonical FSM we recovered after running the extraction procedure in Section VI. There is not a one-to-one mapping between the intermediary representation extracted from the text and the resulting state machines for four reasons: 1) Not all Canonical FSM behaviors are clearly and unambiguously described in the text. 2) Some behaviors are mentioned more than once, giving us several opportunities to extract an expected transition. 3) We have annotated for a larger set of behaviors than needed to extract the communication transitions, we do this to be able to capture the language used to express FSM behaviors. 4) The metrics shown in Tables I and II are based on text span matching, however, we do not need to have a strict match in a text segment to successfully recover a behavior.

Our results show that learning technical word representations is useful for the task of extracting FSM information from protocol specifications. We demonstrate that we can recover a significant portion of the \xmlrep~for the six evaluated protocols. Moreover, we show that we can recover partially correct FSMs by using the procedure outlined in Section~\ref{sec:fsmextraction}. This analysis indicates that the grammar proposed in Section~\ref{sec:annotation} can capture enough information to reconstruct a significant portion of the FSM, while being general enough to be applied to various protocols.
Ambiguity and missed information in the RFCs result in transitions being partially/incorrectly recovered or missed. We show examples in the Appendix and discuss limitations in Section \ref{sec:disc}.

\subsection{Attacker Synthesis Evaluation}
\label{sec:eval_synthesis}

In this section we use \textsc{Korg}~\cite{korg} 
to automatically synthesize attackers against
the TCP 
and DCCP
connection establishment and tear-down routines. 
Note we cannot extract Canonical FSMs like the ones manually derived and used by~\cite{korg}. Our FSMs are partial, and we had to modify \textsc{Korg} to make it  work with partial FSMs. We also had to modify \textsc{Korg} to support DCCP. We use our modified-\textsc{Korg} on all the models including the Canonical FSM and report these results below.

\subsubsection{Methodology}
We apply the same methodology to TCP and DCCP.
We use the \xmlrep s obtained with the models with best results for transition extraction (\linearrules~and~\bertrules), and Gold.
We then extract \fsm s and transpile them  to \program s.
All \fsm s are presented in Appendix \ref{sec:fsm}.

We synthesize attackers that invalidate the properties from Eqn.~\ref{eqn::tcp_props} for TCP 
and Eqn.~\ref{eqn::dccp_props} for DCCP. Given a property and a 
{\program}, we can only use \textsc{Korg} if the program supports
	the property. 
We check what properties are supported by each program and present the results
 in Table~\ref{table::fsm_sups}.

We ask \textsc{Korg} to synthesize at most 100 
attackers which we refer to as {\em candidate attackers} 
because they might not work against the protocol's Canonical \program.  
We check the {\em candidate attackers} 
against the corresponding Canonical \program; those that succeed are {\em confirmed attackers}. Unconfirmed attackers can be thought of as false positives.

\begin{table}[H]
\centering
\caption{Properties Supported by Each \textsc{Promela} Program (checkmark/x-mark means property is supported/not supported).}
\begin{tabular}{|l|llll|}
\hline
TCP \program          
             & $\models \phi_1$ & 
               $\models \phi_2$ & 
               $\models \phi_3$ & 
               $\models \phi_4$ \\\hline
Canonical    & \cmark & 
               \cmark &
               \cmark &
               \cmark \\
Gold         & \cmark &  
               \cmark &  
               \xmark &  
               \xmark \\
\linearrules   
             & \cmark &
			\cmark &
			\xmark &
			\cmark \\
\bertrules
             & \cmark &
			\cmark &
               \xmark &
               \cmark\\\hline
DCCP \program          
             & $\models \theta_1$ & 
               $\models \theta_2$ & 
               $\models \theta_3$ & 
               $\models \theta_4$ \\\hline
Canonical   
             & \cmark & 
               \cmark &
               \cmark &
               \cmark \\
Gold         
             & \cmark & 
               \cmark &
               \cmark &
               \cmark \\
\linearrules      
             & \cmark & 
               \cmark &
               \cmark &
               \cmark \\
\bertrules        
             & \cmark & 
               \cmark &
               \cmark &
               \cmark \\ \hline              
\end{tabular}
\label{table::fsm_sups}
\end{table}

\subsubsection{Supported Properties}

{\em Why do 
noisier models for TCP support a property the Gold model does not support?}
As shown in Table~\ref{table::fsm_sups}, the TCP Gold {\program} does 
not support property~$\phi_4$, while the TCP {\linearrules} and {\bertrules}
{\program}s do.  This might seem counterintuitive, as the Gold \program~is derived from the Gold \xmlrep, which is theoretically less noisy than the {\linearrules} and {\bertrules} {\xmlrep}s. Recall that $\phi_4$ relates to connection tear-down from the TCP state \texttt{SYN\_RECEIVED}. Upon investigation, we found that the TCP Gold {\program} violates $\phi_4$ because of a single erroneous transition from \texttt{SYN\_RECEIVED} to \texttt{CLOSE\_WAIT}, and a missing \texttt{SYN?} event in the transition from \texttt{SYN\_SENT} to \texttt{SYN\_RECEIVED}.  While the TCP {\linearrules} and {\bertrules} {\program}s contain similar erroneous transitions from \texttt{SYN\_RECEIVED}, they nonetheless satisfy $\phi_4$ because their erroneous transitions are never enabled. Basically, the same erroneous transition manifests in all three TCP {\program}s, but in the TCP Gold {\program} the code is reachable, while in the TCP {\linearrules} and {\bertrules} {\program}s it is unreachable.

{\em Why do TCP and DCCP have such different support for
properties intended to capture comparable behavior?}
In Table~\ref{table::fsm_sups}, we notice that the TCP Gold, {\linearrules}, and {\bertrules}
{\program}s all violate $\phi_3$, meaning they all have stuck states.
For DCCP, all {\program}s support~$\theta_1$ and~$\theta_3$, meaning they never self-loop into a stuck state, or self-loop forever.  Notably, either case would constitute a stuck state.  It seems strange that the TCP {\program}s would be so susceptible to stuck states, while the DCCP {\program}s are apparently invulnerable to a closely related problem.
Further investigation revealed that
in contrast to TCP, DCCP does not support active/active establishment.
Hence in order for a DCCP \program~to support connection establishment, it requires both an active and a (matching) passive
establishment routine.
The DCCP Gold, {\linearrules}, and {\bertrules} {\program}s
all capture the active establishment routine but not the passive one.
Therefore, in all three \program s, none of the states containing self-loops are reachable,
and so~$\theta_1$ and~$\theta_3$ are vacuously supported.

\subsubsection{Examples of Attacks}
Table~\ref{table::attack_transfers} presents 
the {\em candidate attackers} generated 
for all programs and properties and false positives. 
We present some examples of {\em confirmed attackers}.
Each example $A$ is named following the convention $\textit{protocol}.M.\alpha.N$, 
	where \emph{protocol} is TCP or DCCP, and
	$A$ was the $N^{\text{th}}$ \program~output by \textsc{Korg}
	when given the \emph{protocol} \program~$M$ and property $\alpha$.

\begin{itemize}
	\item \emph{TCP.\bertrules.$\phi_1$.32} injects a single \texttt{ACK} to Peer 2, causing a desynchronization between the peers which can eventually cause a half-open connection, violating~$\phi_1$.
	\item \emph{DCCP.\linearrules.$\theta_4$.32} injects and drops messages to and from each peer
	to first (unnecessarily) start and abort numerous connection routines,
		then guide both peers at once into \texttt{CLOSE\_REQ}, violating $\theta_4$.
	\item \emph{DCCP.\bertrules.$\theta_2$.96} is programmatically different from
	\emph{DCCP.\linearrules.$\theta_4$.32}, but violates $\theta_4$ using basically the
        same approach.
\end{itemize}

\begin{table}[H]
\newlength{\width}
\width3mm
\centering
\setlength\tabcolsep{3.5pt}
\caption{Candidate and Unconfirmed Attacks Synthesized using each \textsc{Promela} Program $P$ and Correctness Property $\varphi$. If P does not support $\varphi$, \textsc{Korg} cannot generate any attackers.}
\begin{tabular}{|p{9.9\width}|p{0.9\width}p{0.9\width}p{0.9\width}p{0.9\width}|p{0.9\width}p{0.9\width}p{0.9\width}p{0.9\width}|}
\hline  & \multicolumn{4}{p{3\width}|}{
						Candidates Guided~by~$\varphi$.
					} 
        & \multicolumn{4}{p{5\width}|}{
        		Unconfirmed Candidates Guided~by~$\varphi$.} \\\hline
TCP \program	          
             & $\phi_1$ & 
               $\phi_2$ & 
               $\phi_3$ & 
               $\phi_4$ 
             & $\phi_1$ & 
               $\phi_2$ & 
               $\phi_3$ & 
               $\phi_4$\\\hline
Canonical & 1 & 
            9 & 
            36 & 
            17 & 
            0 & 
            0 & 
            0 & 
            0 \\
Gold    
		   & 2 & 
               0 & 
               0 & 
               0    
             & 0 & 
               0 & 
               0 & 
               0 \\
\linearrules
		   & 1 & 
               0 & 
               0 & 
               0   
             & 0 & 
               0 & 
               0 & 
               0 \\
\bertrules   
             & 1 & 
               0 &
               0 & 
               0    
             & 0 & 
               0 & 
               0 & 
               0 \\\hline
DCCP \program          
             & $\theta_1$ & 
               $\theta_2$ & 
               $\theta_3$ & 
               $\theta_4$ 
             & $\theta_1$ & 
               $\theta_2$ & 
               $\theta_3$ & 
               $\theta_4$ \\\hline
Canonical & 0 & 
            12 & 
             0 & 
             1 & 
             0 & 
             0 & 
             0 & 
             0 \\
Gold
             & 0 & 
               1 & 
               0 & 
               1         
             & 0 & 
               0 & 
               0 & 
               0 \\
\linearrules
             & 8 &
               2  &
               13 &
               1    
             & 2 & 
               0 & 
               13 & 
               0 \\
\bertrules 
             & 5 & 
               2 & 
               9 & 
               1       
             & 2 & 
               0 & 
               9 & 
               0 \\\hline          
\end{tabular}
\label{table::attack_transfers}
\end{table}

\subsubsection{Candidate Attackers}   

\emph{Why does property $\phi_2$ not yield candidate attackers with TCP?}
In detail, $\phi_2$ says: ``if the two peers infinitely often revisit the configuration where the first is in \texttt{LISTEN} while the second is in \texttt{SYN\_SENT}, then eventually the first peer will reach \texttt{ESTABLISHED}''.
In the TCP Gold, \linearrules, and \bertrules~\program s, 
	the tear-down routine is incomplete, so a connection cannot be
	fully closed.  Moreover, the \emph{timeout} transitions needed to abort
	a connection establishment are missing.
Hence these \program s~cannot capture the antecedent of $\phi_2$,
	where two peers ``infinitely often revisit the configuration where the first is in \texttt{LISTEN} while the second is in \texttt{SYN\_SENT}''.
Since the \program s satisfy $\phi_2$ only vacuously, they cannot be used 
by \textsc{Korg} to generate {\em candidate attackers} with $\phi_2$.

\emph{Why does property $\phi_4$ not yield candidate attackers with the TCP \linearrules~or \bertrules~\program s?}
In the TCP \linearrules~and \bertrules~\program s, \texttt{SYN\_RECEIVED} is unreachable because of two missing transitions.  Therefore, the TCP \linearrules~and \bertrules~\program s support $\phi_4$ only vacuously and thus cannot be used with \textsc{Korg} to generate any candidate attackers using $\phi_4$.
Either of the missing transitions would fix the problem 
	(the TCP Gold \program~has one).

\emph{Why does property $\theta_3$ not yield confirmed attackers with DCCP?}
As shown in Table~\ref{table::attack_transfers}
none of the candidate DCCP attackers generated using property $\theta_3$ are confirmed.
We investigated and found that for the canonical model 
 the attacker can not violate $\theta_3$,
	unless it is allowed to loop forever, i.e. the attack is continuous, a 
	different (and less realistic) attacker model than the one we consider.

\subsubsection{Comparison to Canonical Attacker Synthesis}
For each attack synthesized using the TCP Gold, \linearrules, or \bertrules~FSM, a similar attack was also synthesized using the TCP Canonical FSM.  However, attacks found using TCP Canonical FSM exhibited five overarching strategies, of which attacks found using TCP Gold, \linearrules, or \bertrules~FSM, exhibited only one.  

Using the DCCP Gold, \linearrules, or \bertrules~FSM, we find numerous attacks all of which passively spoof both peers in order to guide the peers into \texttt{CLOSE\_REQ}$\times$\texttt{CLOSE\_REQ}.  We cannot find active-spoofing attacks using the DCCP Gold, \linearrules, or \bertrules~FSM, because these FSMs lack a functional passive establishment routine for active-spoofing to interact with.  In contrast, all of the DCCP Canonical attacks use active spoofing.  DCCP Canonical has both active and passive establishment, but in this case the \textsc{Spin} model-checker finds counter-examples where the peers do passive establishment first.

We show examples of attacks synthesized with the canonical FSM, but
not with the NLP generated FSMs in the Appendix.

\subsubsection{Summary}

Our NLP pipeline and attacker synthesis task successfully generated
several confirmed attackers against two representative protocols: TCP and DCCP.
However, our method depends on the accuracy of the NLP extraction task, 
the correctness of the extracted FSM,
the quality of the selected properties, and the power of the attack synthesis tool. 
We discuss limitations and improvement directions in Section~\ref{sec:disc}.

%% file: related.tex

\section{Related Work}
\label{sec:rel}
Below we present related works across three categories.

{\em Logical Information Extraction.}
Rule-based systems like WHYPER~\cite{whyper} and DASE~\cite{dase} identify sentences describing mobile application permissions and extract command-line input constraints from manual pages, respectively. Witte \etal~\cite{TM} use rules over documentation and source code to create an ontology allowing the cross-linking of software artifacts.

Other works combine NLP with techniques from traditional software engineering and security. Lin \etal~\cite{lin2008automatic} infer protocol formats by combining NLP with program analysis. NLP has also been used to gather \emph{threat intelligence} by interacting with botnets~\cite{small2008catch}, logically contrasting CVEs~\cite{dong2019towards}, or analyzing bug reports in the context of  data collected with a honeypot~\cite{feng2019understanding}.

Ding and Hu~\cite{10.1145/3243734.3243865} used pre-trained word embeddings to identify physical channels in IoT from application descriptions. Tian \etal~\cite{203866} used pre-trained word vectors and other standard NLP features to compare security policy descriptions written in text in the context of IoT application authorization. Both works relied on off-the-shelf NLP tools,  and worked over keywords in isolation, or over short and simple sentences.

Recently, Jero \etal~\cite{JPGN_iaai_2019} proposed a system to extract protocol rules from textual specifications for grammar-based fuzzing. They also took a zero-shot learning approach to adapt to protocols that are unseen at training time. However, they focused on a limited set of properties, and did not explicitly model the behavior of the protocol. More closely related to our work, Chen \etal~\cite{chen2019devils} explored the use of NLP to discover logical vulnerabilities in payment services. They extended the FSMs for evaluated payment services by using the dependency parse tree of sentences in a developer guide to extract the parties involved in the process, as well as the content transmitted between them. To identify relevant sentences, they used word embeddings trained on relevant documentation. In our work, we also leverage word representations trained on in-domain data. However, we aim to reconstruct the full FSM from the text, while they relied on a manually implemented FSM. While their language analysis was done at the sentence-level, we predict logical flow structures that span multiple sentences.

{\em Full Correctness Specification.} 
Zhai \etal~\cite{zhai2020c2s} 
	 automatically extract formal software specifications from comments in the implementation code. Zhang \etal~\cite{zhang2020automated} use NLP to extract LTL correctness specifications from prose policies for IoT apps.
In contrast to our work, they assume that the actual software code is known ahead of time. Other related works infer abstract protocol implementations using network traces~\cite{comparetti2009prospex,wang2011inferring,cho2010inference}, program analysis~\cite{cho2011mace}, or model checking~\cite{lie2001simple,corbett2000bandera}. These approaches rely extensively on input from human experts and do not easily generalize to new software or protocols.

{\em Implementation Extraction.} Yen \etal~\cite{nlp_sigcomm_2022} explored the use of NLP techniques to map RFCs to protocol implementations. To do this, they manually engineer an existing semantic parser to handle networking-specific vocabulary, and translate individual sentences to logical forms that can then be mapped to executable functions. They include the spec author in the loop to disambiguate cases where the functionality is under-specified. 
They do not perform any task-specific learning, and they work at the sentence-level.

%% file: discussion.tex

\section{Limitations}
\label{sec:disc}

In this section we discuss some of the limitations of our approach
and directions for improvement.

\textbf{Why our NLP models could not extract Canonical FSMs from RFCs.} 
Canonical FSMs are created based not
only on RFCs but also on input from experts with exposure to protocol
implementations, and often also rely on analyzing the code \mbox{\cite{abbr_raid_2020,quic_sp_2015,quic_sigcomm_2019}}. 
RFCs contain ambiguities, unspecified behaviors that human experts solve in creating the Canonical FSM \mbox{\cite{jero2015leveraging,nlp_sigcomm_2022}}, or simply missing information. Thus, unlike traditional NLP semantic parsing problems~\mbox{\cite{Kate2005LearningTT,kwiatkowksi-etal-2010-inducing,cheng-etal-2017-learning}}, which study methods for translating natural language into a complete formal representation, in our setting there is not a complete one-to-one translation between the text and the FSM.  We address this challenge by defining an intermediary semantic representation that can be extracted unambiguously from the text, and then use this intermediary representation as the basis for the FSM extraction. The ground truth for these intermediary representations is what we refer to as Gold intermediary representations.

One avenue to extract better FSMs, possibly canonical ones, is to solve ambiguities existing in the text by leveraging human expertise. This can be done by using NLP methods that exploit unlabeled data and human knowledge. A potential direction for improvement is to design learning objectives that, in addition to exploiting domain-specific corpora, can augment the intermediary representations and constraint the predictions using structured domain knowledge.

 \textbf{Limitations of attacker synthesis with partial FSMs.}
 The partial FSMs produced by the NLP pipeline combined with the FSM extraction algorithm exhibit numerous errors, which impacted our ability to use these FSMs for attacker synthesis.  Some attacks which could be found using the Canonical FSMs were not found using the partial FSMs, and, some of the attacks found using the partial FSMs were not confirmed on the Canonical FSMs. There are two causes for these mistakes: missed transitions and incorrect transitions.

 One direction to address these limitations is by leveraging protocol completion \mbox{\cite{alur2017automatic}}, where given an incomplete protocol FSM and some properties, the goal is to strategically add transitions so that the completed FSM supports all the properties.  Their solution relied on counterexample-guided inductive synthesis (CEGIS) \mbox{\cite{alur2014synthesizing}}.  Our problem is a little more difficult, because in addition to missing transitions, we also need to worry about incorrect transitions, so the approach used in
 \mbox{\cite{alur2017automatic}} would need to be modified such that the solver is also allowed to delete or edit transitions.  Another approach would be to leverage prior work in automatic program repair \mbox{\cite{bonakdarpour2012automated}.}

 \textbf{Selecting properties.}
 The attackers we find are driven by the selection of properties that 
 the Canonical and extracted \fsm s support.
For attacker synthesis, the most useful properties 
    describe critical functionality of a protocol,
    for example, that it must reliably open and close connections,
    or that it must not deadlock.
We also prefer properties that are not too implementation-specific,
    because there are multiple ways to implement a protocol
    while still achieving the intended functionality,
    as illustrated for Alternating Bit Protocol in~\mbox{\cite{alur2017automatic}}. 

Protocol correctness properties should be provided by protocol
designers. Unfortunately, protocols are often implemented and deployed before
 textual specifications are published. 
This is 
the case with QUIC, which was deployed without detailed public specification
or analysis. The authors of a 2015 QUIC security analysis~\mbox{\cite{quic_sp_2015}} mention that they
relied on code and discussion with protocol developers to derive a
protocol description as the available documentation was insufficient.

\textbf{Extracting properties.}
While several NLP works looked at converting natural language statements into properties expressed in temporal logic, RFCs do not have a dedicated section detailing protocol correctness properties in an explicit and succinct way. Instead, humans identify these properties by observing the behaviors emerging from the specification and inferring the intent behind them, or by reading prose descriptions of the developer's intention. One promising approach is to study these inference processes and formulate them as NLP problems that take into account the functionality described by the protocol as part of the input. Rather than converting the explicit textual statement into properties, one can define an abductive process that infers relevant desired properties of the extracted model  and rely on textual description of protocol tests for specifications that offer similar functionality.

\textbf{Limitations of \textsc{Korg}.}
\textsc{Korg} was not designed for broken or partial \fsm s (expressed as \program s),
    that might violate or vacuously satisfy the provided properties.  
    In these cases it might generate no candidates whatsoever,
        or some candidates, none of which are confirmed.
Also, \textsc{Korg} outputs many identical or similar candidates,
    but we would prefer a diversity of candidate attackers so that if
    some are not confirmed, perhaps others will be.
The problem of determining when two candidate attackers are
    similar reduces to defining an equivalence relation on counterexamples,
    as studied in~\mbox{\cite{vick2021counterexample}}.  
Perhaps such work could be leveraged to quotient \textsc{Korg}'s 
search-space by the equivalence class of the candidates it already found,
    resulting in a diversity of attackers.

\textbf{Generalizability to other RFCs.} 
While we consider a set of 6 different protocols, including TCP (one of the most
well-known and used protocols), there are further aspects we did not consider
in this work. One such aspect is considering changes in RFCs.
We believe that one promising direction for investigating changes in RFCs and
impact on FSMs is investigating congestion control protocols that share a common 
approach in detecting congestion, where newer refinements were proposed 
enhancing the original protocol. We expect that while we can use the same
technical domain knowledge we might need to update our grammar to handle changes. 

We did not consider secure protocols in this work. Note that QUIC
was just recently standardized in May 2021, as RFC 9000~\mbox{\cite{rfc9000}}. 
Here we can focus on RFC drafts changes for QUIC and TLS 1.3, 
particularly the key exchange aspects. Secure protocols will most likely 
require us to refine both the grammar and the domain knowledge  we 
built for this work.

%% file: extraction_pseudocode.tex
\begin{tcolorbox}[
	colback=white,
	colframe=black,
	sharp corners,
	title={Alg.~\textsc{extractTran}$(\textsf{xml}, \textsf{T})$}]

\textbf{Inputs:}
\begin{itemize}
	\item \textsf{xml} \xmlrep
	\item \texttt{transition} block \textsf{T}, contained with \textsf{xml}.
\end{itemize}

\textbf{Outputs:}
\begin{itemize}
	\item A set $T_{\textsf{T}}$ containing potential transitions $s \xrightarrow[]{\ell} s'$ described in and around the block \textsf{T}.
\end{itemize}

\par\noindent\rule{\textwidth}{0.4pt}

\begin{enumerate}[1.]
	\item $\emph{from} := \textsc{extractSourceState}(\textsf{T}, \textsf{xml})$
	\item $\emph{to} := \textsc{extractTargetState}(\textsf{T}, \textsf{xml})$
	\item $\emph{int} := \textsc{extractIntermediaryStates}(\textsf{T}, \textsf{xml})$
	\item $\textsf{C} := \textsc{closestControlContaining}(\textsf{T}, \textsf{xml})$
	\item $\emph{outer} := [ \, ]$
	\item If $(\emph{to} = \emph{null} \text{ and } \emph{from} = \emph{null})$:
	\begin{enumerate}[1.]
		\item $\emph{to} := \textsc{scanChildrenForTargetState}(\textsf{T})$
	\end{enumerate}
	\item If $(\emph{to} = \emph{null} \text{ or } \emph{from} = \emph{null})$:
	\begin{enumerate}[1.]
		\item $\emph{outer} := \textsc{scanContextForStates}(\textsf{C}, \textsf{T})$
	\end{enumerate}
	\item $\ell := \epsilon$
	\item $i := 1$
	\item While (not $\textsc{searchedEnough}(\ell, \emph{outer}, i, \textsf{or}, \textsf{C})$):
	\begin{enumerate}[1.]
		\item If $\ell = \epsilon$:\\
		\texttt{/*} $\ell$ is the transition label,
		\textsf{brk} indicates if the source states are given outside \textsf{C}, and \textsf{or} indicates if $\ell$ is of the form ``$\ell_0$ or $\ell_1$ or ... or $\ell_k$''. \texttt{*/}
		\begin{enumerate}[1.]
			\item $(\ell, \textsf{brk}, \textsf{or}):=\textsc{extractTranLbl}(\textsf{T}, \textsf{C})$.
		\end{enumerate}
		\item If $\emph{outer} = [ \, ]$ and $(\emph{from} = \emph{null}$ or $\emph{to} = \emph{null})$:
		\begin{enumerate}[1.]
			\item $\emph{outer} := \textsc{scanContextForStates}(\textsf{C}, \textsf{T})$
		\end{enumerate}
		\item $\textsf{C} := \textsc{closestControlContaining}(\textsf{C}, \textsf{xml})$
		\item $i++$
	\end{enumerate}
	\item $(\emph{fromS}, \emph{to}) := \textsc{fixFromToStates}(\emph{from}, \emph{to}, \emph{outer})$
	\item If $\emph{int} \neq [\,]$:
	\begin{enumerate}[1.]
		\item $(\ell_0, ..., \ell_j) := \textsc{partitionLabelAcross}(\ell, \emph{int})$
		\item Let $S_0 := \{ s_0 \xrightarrow[]{\ell_0} s_1 \mid s_0 \in \emph{fromS} \}$
		\item Let $S_1 := \{ s_1 \xrightarrow[]{\ell_1} s_2, ..., s_j \xrightarrow[]{\ell_j} \emph{to} \}$
		\item Return $S_0 \cup S_1$
	\end{enumerate}
	\item If $\textsf{brk} = \emph{true}$:
	\begin{enumerate}
		\item $\textsf{C} := \textsc{closestControlContaining}(\textsf{T}, \textsf{xml})$
		\item $\textsf{C}' := \textsc{closestControlContaining}(\textsf{C}, \textsf{xml})$
		\item $\textsf{fromS} := \textsc{scanContextForStates}(\textsf{C}', \textsf{C})$
	\end{enumerate}
	\item Return $\{ s_0 \xrightarrow[]{\ell} \emph{to} \mid s_0 \in \emph{fromS} \}$.
\end{enumerate}

\end{tcolorbox}

%% file: grammar.tex
\subsection{Grammar Examples}
\label{app:grammarexample}

Figure \ref{fig:ann_example} shows an example of an annotated block from the TCP RFC. Here, we can observe a list of events within one control statement.

\input{flow_control_tags_example}

%% file: flow_control_tags_example.tex
\begin{figure}[ht]
\begin{lstlisting}
<control>
	<trigger>
		if active and the foreign socket is
		specified,
	</trigger>
	<action type="issue">
		issue <arg>a <ref_event id="10">SYN</ref_event> segment</arg>.
	</action>
	<variable>
		An initial send sequence number (ISS) is
		selected.
	</variable>
	<action type="send">
		A <arg><ref_event id="10">SYN</ref_event> segment of the form 
		<SEQ=ISS><CTL=SYN></arg> is sent.
	</action>
	<variable>
		Set SND.UNA to ISS, SND.NXT to ISS+1,
	</variable>
	<transition>
		enter the <arg_target><ref_state id="2">SYN-SENT</ref_state> state<arg_target>
	</transition>
</control>
\end{lstlisting}
\caption{Example of flow control annotations for TCP.}\label{fig:ann_example}
\end{figure}

%% file: segmentation.tex
\subsection{Segmentation Results}
In Table \ref{tab:segmentation_strategies}, we show the detailed performance of different segmentation strategies to create the base textual unit in our sequence-to-sequence models. 

\begin{table*}[ht]
\centering
      \caption{Average Results for Different Segmentation Strategies (\linear)}\label{tab:segmentation_strategies}
	\begin{tabular}{| l | c | c |  c | c | c | c | c |}
		\hline
		\multirow{2}{*}{Segmentation} & \multicolumn{3}{|c|}{Token-level} & \multicolumn{4}{c|}{Span-level} \\ \cline{2-8}
~ &		Acc & Weighted F1 & Macro F1 &  Strict & Exact & Partial & Type \\ \hline
Token  & 60.37 
            & 59.58 
		   & 44.76 
		   & 31.36 
		   & 36.14 
		   & 59.78 
		   & 58.81 \\
Chunk  & \textbf{62.02}  
		   & \textbf{61.25} 
		   & 46.36 
		   & 33.48 
		   & 39.11 
		   & 62.19 
		   & 62.14 \\
Phrase & 58.95
		   & 56.61 
		   & \textbf{49.58} 
		   & \textbf{63.98} 
		   & \textbf{85.65}
		   & \textbf{85.65}
		   & \textbf{63.98} \\
\hline
	\end{tabular}
\end{table*}

%% file: fsmerrors.tex
\subsection{FSM Extraction Errors Examples}
In Table \ref{tab:fsm_errors}, we show examples of FSM extraction errors.

\begin{table*}[ht]
\centering
 \caption{Examples of FSM Extraction Errors}\label{tab:fsm_errors}
\begin{tabular}{|p{0.15\textwidth}|l|p{0.1\textwidth}|p{0.1\textwidth}|p{0.22\textwidth}|}
\hline
FSM & Transition & Error Type & Reason & Text Excerpt \\
\hline 
Gold TCP & \texttt{FIN\_WAIT\_1}  $\xrightarrow{\texttt{FIN!}}$ \texttt{LAST\_ACK} & Not Found  & Target state not explicit &  CLOSE-WAIT STATE: Since the remote side has already sent FIN, RECEIVEs must be satisfied by text already on hand, but not yet delivered to the user. \\
\hline
Gold DCCP & \texttt{PARTOPEN} $\xrightarrow{\texttt{DCCP-CLOSE?}}$ \texttt{OPEN} & Incorrect & Text is ambiguous & The client leaves the PARTOPEN state for OPEN when it receives a valid packet other than DCCP-Response, DCCP-Reset, or DCCP-Sync from the server. \\
\hline
\linearrules~and \bertrules & \texttt{SYN\_SENT} $\xrightarrow{\texttt{SYN!ACK!}}$ \texttt{SYN\_RECEIVED}  & Partially Recovered (expected \texttt{SYN?ACK!}) & Receive action is not explicit & If the state is SYN-SENT then enter SYN-RECEIVED, form a SYN,ACK segment and send it. \\
\hline
\end{tabular}
\end{table*}

%% file: fsmfigures.tex

\subsection{Finite State Machine Figures}
\label{sec:fsm}

We present FSMs for TCP and DCCP in Figures \ref{appendix:fsm:figures:1},\ref{appendix:fsm:figures:tcp:bert:linear} and
\ref{appendix:fsm:figures:canonical}.
Note that in the DCCP diagrams we omit the states CHANGING, STABLE, and UNSTABLE, which are described in the RFC but are (a) unreachable dead code in all the extracted FSMs and (b) unrelated to the connection routine.  We use $*$ as a wild-card, $!$ to mean \emph{send}, $?$ to mean \emph{receive}, \texttt{==} to denote variable-\emph{reading}, and \texttt{:=} to denote variable-\emph{writing}.

\begin{figure*}[h!]
\begin{subfigure}{.5\textwidth}
  \include{fsm-figures/dccp/golden/golden.dccp}
\end{subfigure}
\begin{subfigure}{.5\textwidth}
  \include{fsm-figures/dccp/bert/bert.dccp}
\end{subfigure}
\begin{subfigure}{.6\textwidth}
  \include{fsm-figures/dccp/linear/linear.dccp}
\end{subfigure}
\begin{subfigure}{.4\textwidth}
  \include{fsm-figures/tcp/golden/golden.tcp}
\end{subfigure}
\caption{DCCP Gold, \bertrules, and \linearrules~\fsm s; and TCP Gold \fsm.}
\label{appendix:fsm:figures:1}
\end{figure*}

\begin{figure*}[h!]
  \include{fsm-figures/tcp/bert/bert.tcp}
\caption{\textbf{TCP \bertrules~and \linearrules~\fsm}.  (They are identical.)}
\label{appendix:fsm:figures:tcp:bert:linear}
\end{figure*}

\begin{figure}[h!]
\begin{subfigure}{.45\textwidth}
  \include{fsm-figures/tcp/tcp}
\end{subfigure}
\begin{subfigure}{.45\textwidth}
  \include{fsm-figures/dccp/dccp}
\end{subfigure}
\caption{TCP and DCCP Canonical \fsm s.}
\label{appendix:fsm:figures:canonical}
\end{figure}

%% file: fsm-figures/dccp/golden/golden.dccp.tex
\adjustbox{max width=\textwidth}{
  \centering
  \begin{tikzpicture}

    \node[state, initial]               (CLOSED)   {\scriptsize CLOSED};
    \node[state, below left of=CLOSED, node distance=2cm]  (LISTEN)   {\scriptsize LISTEN};
    \node[state, below right of=CLOSED, node distance=2cm] (REQUEST)  {\scriptsize REQUEST};
    \node[state, below of=LISTEN, node distance=2cm]       (RESPOND)  {\scriptsize RESPOND};
    \node[state, below of=REQUEST, node distance=2cm]      (PARTOPEN) {\scriptsize PARTOPEN};
    \node[state, below of=RESPOND, node distance=2cm]      (OPEN)     {\scriptsize OPEN};
    \node[state, below of=PARTOPEN, node distance=2cm]     (CLOSING)  {\scriptsize CLOSING};
    \node[state, below of=OPEN, node distance=2cm]         (CLOSEREQ) {\scriptsize CLOSEREQ};
    \node[state, below of=CLOSING, node distance=2cm]      (TIMEWAIT) {\scriptsize TIMEWAIT};

    \draw[->] (CLOSED) edge[right] node{\scriptsize \texttt{DCCP\_REQUEST!}} (REQUEST);

    \draw (CLOSEREQ.west) -| node[above right,text width=2cm,xshift=0.1cm]
                                  {\scriptsize \texttt{DCCP\_CLOSE?}
                                               \texttt{DCCP\_RESET!}}
          (-5,-5.5);

    \draw (-5,-5.5) -- 
          (-5,-2);
        
    \draw[-stealth] (-5,-2) to (CLOSED.south west);

    \draw[->] (CLOSEREQ) edge[above left] 
    node[rotate=35,xshift=1.2cm,yshift=0.1cm]{\scriptsize \texttt{DCCP\_CLOSEREQ?}} (CLOSING);

    \draw[->] (CLOSING) edge (TIMEWAIT);

    \draw[->] (LISTEN) edge[left] node[text width=1.9cm,xshift=-0.1cm]
                       {\scriptsize \texttt{DCCP\_REQUEST?}
                                    \texttt{DCCP\_RESPONSE!}}
              (RESPOND);

    \draw[->] (OPEN) edge[left] node{\scriptsize \texttt{DCCP\_CLOSEREQ!}} (CLOSEREQ);
    \draw[->] (OPEN) edge (CLOSING);

    \draw[->] (PARTOPEN) edge[right, loop below]
                         node[text width=1cm,xshift=0.8cm,yshift=0.6cm]
                         {\scriptsize \texttt{DCCP\_DATAACK!}~or
                              (\texttt{DCCP\_RESPONSE? DCCP\_ACK!})} (PARTOPEN);
    \draw[->] (PARTOPEN) edge[bend left=10] (CLOSED);
    \draw[->] (PARTOPEN.south west) to 
                        node[above left]
                         {\scriptsize *\texttt{?}}
                          (OPEN.north east);

    \draw[->] (REQUEST) edge[right] node[text width=1cm]{\scriptsize \texttt{DCCP\_RESPONSE?}
                                                     \texttt{DCCP\_ACK!}}
               (PARTOPEN);

    \draw[->] (RESPOND) edge[left] node{\scriptsize \texttt{DCCP\_ACK?}} (OPEN);
    \draw[->] (RESPOND) edge[above left,bend right] 
    node[rotate=65,xshift=0.8cm]{\scriptsize \texttt{DCCP\_RESET!}} (CLOSED);

    \draw (TIMEWAIT) to
              ([yshift=-0.8cm]TIMEWAIT)
                         -| 
              (-5,-5.5);
  \end{tikzpicture}}
  \caption{\textbf{DCCP Gold FSM}.}
  \label{fig:dccp.golden}

%% file: fsm-figures/dccp/bert/bert.dccp.tex
\adjustbox{max width=\textwidth}{
  \centering
  \begin{tikzpicture}

    \node[state, initial]               (CLOSED)   {\scriptsize CLOSED};
    \node[state, below left=1cm and 0.5cm of CLOSED]  (LISTEN)   {\scriptsize LISTEN};
    \node[state, below right=1cm and 0.5cm of CLOSED] (REQUEST)  {\scriptsize REQUEST};
    \node[state, below=1cm of LISTEN]       (RESPOND)  {\scriptsize RESPOND};
    \node[state, below=1cm of REQUEST]      (PARTOPEN) {\scriptsize PARTOPEN};
    \node[state, below=1cm of RESPOND]      (OPEN)     {\scriptsize OPEN};
    \node[state, below=1cm of PARTOPEN]     (CLOSING)  {\scriptsize CLOSING};
    \node[state, below=1cm of OPEN]         (CLOSEREQ) {\scriptsize CLOSEREQ};
    \node[state, below=1cm of CLOSING]      (TIMEWAIT) {\scriptsize TIMEWAIT};

    \draw[->] (CLOSED) edge[right] node{\scriptsize \texttt{DCCP\_REQUEST!}} (REQUEST);

    \draw[->] (CLOSEREQ) to (TIMEWAIT);
    \draw[->] (CLOSEREQ.west) -| node[below right,xshift=-0.3cm]
                                  {\scriptsize \texttt{DCCP\_CLOSEREQ!}}
          (-5,-6) to 
          (-5,-2) to (CLOSED.south west);
    \draw[->] (CLOSEREQ) edge[below] 
    node[rotate=25]{\scriptsize \texttt{DCCP\_CLOSEREQ?}} (CLOSING);

    \draw[->] (CLOSING) to (TIMEWAIT);

    \draw[->] (LISTEN) edge[left] 
                       node[text width=1.5cm,xshift=-0.4cm]
                       {\scriptsize \texttt{DCCP\_REQUEST?}}
              (RESPOND);

    \draw[->] (OPEN) edge[loop left] node[left]{\scriptsize \texttt{DCCP\_DATA?}} (OPEN);
    \draw[->] (OPEN) edge[left] node{\scriptsize \texttt{DCCP\_CLOSEREQ!}} (CLOSEREQ);
    \draw[->] (OPEN) to (CLOSING);

    \draw[->] (PARTOPEN.north east)
              to node[right,text width=1cm]{\scriptsize \texttt{DCCP\_RESPONSE? DCCP\_DATA!}} 
              (REQUEST.south east);
    \draw[->] (PARTOPEN) edge[right]
      node[text width=4.5cm,xshift=0.6cm]{\scriptsize \texttt{DCCP\_DATA?} or
                       (\texttt{DCCP\_RESPONSE? DCCP\_RESET? DCCP\_SYNC?})}
              (OPEN);
    \draw[->] (PARTOPEN) edge[loop right] node[right,text width=1cm]{\scriptsize
      \texttt{DCCP\_ACK! DCCP\_RESET?}} (PARTOPEN);

    \draw[->] (REQUEST.west) to
              ([xshift=-1.2cm]REQUEST.west) |-
              node[above right,text width=2cm]
              {\scriptsize \texttt{DCCP\_RESPONSE?} 
                                     \texttt{DCCP\_ACK!}}
              (PARTOPEN.north west);

    \draw[->] (TIMEWAIT) edge[loop below] 
              node[right,text width=1.5cm,xshift=0.3cm]
              {\scriptsize \texttt{DCCP\_RESET?}, or as an $\epsilon$-transition}
              (TIMEWAIT);

    \draw[->] (RESPOND) edge[right] 
              node[text width=2cm]
              {\scriptsize \texttt{DCCP\_ACK?}}
              (OPEN);
    \draw[->] (RESPOND) -| 
              node[right,rotate=90,yshift=0.2cm,xshift=1cm]{\scriptsize \texttt{DCCP\_RESET!}}
              (CLOSED.south west);

  \end{tikzpicture}}
  \caption{\textbf{DCCP \bertrules~FSM}.}
  \label{fig:dccp.bert}

%% file: fsm-figures/dccp/linear/linear.dccp.tex
\adjustbox{max width=\textwidth}{
  \centering
  \begin{tikzpicture}

    \node[state, initial]               (CLOSED)   {\scriptsize CLOSED};
    \node[state, below left=1cm and 0.5cm of CLOSED]  (LISTEN)   {\scriptsize LISTEN};
    \node[state, below right=1cm and 0.5cm of CLOSED] (REQUEST)  {\scriptsize REQUEST};
    \node[state, below=1cm of LISTEN]       (RESPOND)  {\scriptsize RESPOND};
    \node[state, below=1cm of REQUEST]      (PARTOPEN) {\scriptsize PARTOPEN};
    \node[state, below=0.5cm of RESPOND]      (OPEN)     {\scriptsize OPEN};
    \node[state, below=0.5cm of PARTOPEN]     (CLOSING)  {\scriptsize CLOSING};
    \node[state, below=0.5cm of OPEN]         (CLOSEREQ) {\scriptsize CLOSEREQ};
    \node[state, below=0.5cm of CLOSING]      (TIMEWAIT) {\scriptsize TIMEWAIT};

    \draw[->] (CLOSED) edge[right] node{\scriptsize \texttt{DCCP\_REQUEST!}} (REQUEST);

    \draw[->] (CLOSEREQ) to (TIMEWAIT);
    \draw[->] (CLOSEREQ.west) -| node[below,xshift=0.1cm]
                                  {\scriptsize \texttt{DCCP\_CLOSEREQ!}}
          (-4,-6) to 
          (-4,-2) to (CLOSED.south west);
    \draw[->] (CLOSEREQ) edge[right] 
      node[rotate=20,xshift=-1.2cm,yshift=0.2cm]{\scriptsize \texttt{DCCP\_CLOSEREQ?}} (CLOSING);

    \draw[->] (CLOSING) to (TIMEWAIT);

    \draw[->] (LISTEN) edge[left] 
                       node[text width=1.5cm,xshift=-0.4cm]
                       {\scriptsize \texttt{DCCP\_REQUEST?}}
              (RESPOND);

    \draw[->] (OPEN) edge[left] node{\scriptsize \texttt{DCCP\_CLOSEREQ!}} (CLOSEREQ);
    \draw[->] (OPEN) to (CLOSING);

    \draw[->] (PARTOPEN) 
              edge[right] 
              node[text width=1cm]{\scriptsize \texttt{DCCP\_RESPONSE? DCCP\_DATA!}} 
              (REQUEST);

    \draw[->] (PARTOPEN) edge[below right]
      node[text width=9cm,rotate=20,xshift=-1cm]{\scriptsize \texttt{DCCP\_RESPONSE?}}
              (OPEN);

    \draw[->] (PARTOPEN) edge[loop right] 
    node[below right,text width=1cm,xshift=-0.8cm,yshift=-0.2cm]
    {\scriptsize
      \texttt{DCCP\_ACK! DCCP\_RESET?}} (PARTOPEN);

    \draw[->] (REQUEST.west) to
              node[right,text width=1cm,rotate=85,xshift=-1cm,yshift=0.5cm]
              {\scriptsize \texttt{DCCP\_RESPONSE?} 
                                     \texttt{DCCP\_ACK!}}
              (PARTOPEN.north west);

    \draw[->] (RESPOND) edge[right] 
              node[text width=2cm]
              {\scriptsize \texttt{DCCP\_ACK?}}
              (OPEN);
    \draw[->] (RESPOND) -| 
              node[below]{\scriptsize \texttt{DCCP\_RESET!}}
              (CLOSED);

    \draw[->] (TIMEWAIT) edge[loop below] 
              node[below,text width=1.2cm]
              {\scriptsize \texttt{DCCP\_RESET?}, or as an $\epsilon$-transition}
              (TIMEWAIT);
  \end{tikzpicture}}
  \caption{\textbf{DCCP \linearrules~FSM}.}
  \label{fig:dccp.linear}

%% file: fsm-figures/tcp/golden/golden.tcp.tex
\adjustbox{max width=\textwidth}{
	\centering
	\begin{tikzpicture}
		\node[state, initial] (CLOSED) {\scriptsize CLOSED};
		\node[state, below right=1cm and 0.8cm of CLOSED] (SYN_SENT) {\scriptsize SYN\_SENT};
		\node[state, below=1cm of CLOSED] (LISTEN) {\scriptsize LISTEN};
		\node[state, below left=1cm and 0.5cm of LISTEN] (SYN_RECEIVED) {\scriptsize SYN\_RECEIVED};
		\node[state, below=1cm and 0.5cm of SYN_SENT] (ESTABLISHED) {\scriptsize ESTABLISHED};
		\node[state, below=1cm of SYN_RECEIVED] (FIN_WAIT_1) {\scriptsize FIN\_WAIT\_1};
		\node[state, below=1cm of ESTABLISHED] (CLOSE_WAIT) {\scriptsize CLOSE\_WAIT};
		\node[state, below=1.5cm of CLOSE_WAIT] (CLOSING) {\scriptsize CLOSING};
		\node[state, below right=1cm and 0.5cm of FIN_WAIT_1] (FIN_WAIT_2) {\scriptsize FIN\_WAIT\_2};
		\node[state, below left=1cm of CLOSED] (LAST_ACK) {\scriptsize LAST\_ACK};
		\node[state, below=1.2cm of FIN_WAIT_2] (TIME_WAIT) {\scriptsize TIME\_WAIT};

		\draw[->] (CLOSED) edge[left] node{\scriptsize \texttt{SYN!}} (LISTEN);
		\draw[->] (CLOSED) edge[above right, text width=2cm] node{\scriptsize \texttt{SYN!}} (SYN_SENT);

		\draw[->] (LISTEN) edge[above] node{\scriptsize \texttt{SYN!}} (SYN_SENT);

		\draw[->] (SYN_SENT) edge[below] node[rotate=30]{\scriptsize \texttt{ACK!}} (SYN_RECEIVED);

		\draw[->] (SYN_RECEIVED) edge[above] node{\scriptsize \texttt{ACK?}} (ESTABLISHED);

		\draw[->] (ESTABLISHED)
		          edge node[above,rotate=25,xshift=0.5cm]
		          {\scriptsize \texttt{FIN!}} 
		          (FIN_WAIT_1);
		\draw[->] (ESTABLISHED) edge[right] 
					node[xshift=0.1cm,text width=2cm]{\scriptsize \texttt{FIN? ACK! ACK?}} (CLOSE_WAIT);

		\draw[->] (FIN_WAIT_1) 
		          |- node[above right]{\scriptsize \texttt{ACK?}} 
		          (TIME_WAIT.north west);
		\draw[->] (FIN_WAIT_1) 
		          to 
		          node[left]
		          {\scriptsize \texttt{ACK?}} 
		          (FIN_WAIT_2);

		\draw[->] (CLOSE_WAIT.south east) to[below right] node[rotate=80]{\scriptsize \texttt{FIN!}} (CLOSING.north east);
		\draw[->] (CLOSE_WAIT) edge[loop below] 
			node[xshift=-1.3cm,yshift=0.5cm]{\scriptsize \texttt{FIN? ACK! ACK?}} (CLOSE_WAIT);

		\draw[->] (FIN_WAIT_2) edge[right] node[text width=1cm]{\scriptsize \texttt{FIN? ACK! ACK?}} (TIME_WAIT);

		\draw[->] (CLOSING) |- node[above right]{\scriptsize \texttt{ACK?}} (TIME_WAIT);
		\draw[->] (CLOSING) edge[loop above] 
			node[xshift=-0.5cm]
			{\scriptsize \texttt{FIN? ACK! ACK?}} (CLOSING);

		\draw[->] (LAST_ACK) to node[above,rotate=45]{\scriptsize \texttt{ACK?}} (CLOSED);
		\draw[->] (LAST_ACK) edge[loop below] node[xshift=0.5cm]{\scriptsize \texttt{FIN? ACK!
			ACK?}} (LAST_ACK);

		\draw (TIME_WAIT) edge[loop below] node{\scriptsize \texttt{FIN? ACK!
			ACK?}} (TIME_WAIT);

	\draw[->] (SYN_RECEIVED) 
		          edge[below] 
		          node[rotate=-24,fill=white]{\scriptsize \texttt{FIN? ACK! ACK?}} (CLOSE_WAIT);
	\end{tikzpicture}}
	\caption{\textbf{TCP Gold FSM}.}
	\label{fig:tcp.golden}

%% file: fsm-figures/tcp/bert/bert.tcp.tex
\adjustbox{max width=\textwidth}{
	\centering
	\begin{tikzpicture}
		\node[state, initial] (CLOSED) {\scriptsize CLOSED};
		\node[state, below right of=CLOSED] (SYN_SENT) {\scriptsize SYN\_SENT};
		\node[state, below left of=CLOSED] (LISTEN) {\scriptsize LISTEN};
		\node[state, below=1cm and 0.5cm of LISTEN] (SYN_RECEIVED) {\scriptsize SYN\_RECEIVED};
		\node[state, below=1cm and 0.5cm of SYN_SENT] (ESTABLISHED) {\scriptsize ESTABLISHED};
		\node[state, below left of=ESTABLISHED] (FIN_WAIT_1) {\scriptsize FIN\_WAIT\_1};
		\node[state, below right of=ESTABLISHED] (CLOSE_WAIT) {\scriptsize CLOSE\_WAIT};
		\node[state, below left of=FIN_WAIT_1] (CLOSING) {\scriptsize CLOSING};
		\node[state, below right of=FIN_WAIT_1] (FIN_WAIT_2) {\scriptsize FIN\_WAIT\_2};
		\node[state, below of=CLOSE_WAIT] (LAST_ACK) {\scriptsize LAST\_ACK};
		\node[state, below of=FIN_WAIT_2] (TIME_WAIT) {\scriptsize TIME\_WAIT};

		\draw[->] (CLOSED) edge[left] node{\scriptsize \texttt{SYN!}} (LISTEN);
		\draw[->] (CLOSED) edge[left] node{\scriptsize \texttt{SYN!}} (SYN_SENT);

		\draw[->] (CLOSE_WAIT) edge[above left, near start] node{\scriptsize \texttt{FIN!}} (CLOSING);
		\draw[->] (CLOSE_WAIT) edge[loop below] node{\scriptsize \texttt{FIN? ACK! ACK?}} (CLOSE_WAIT);

		\draw[->] (CLOSING) edge[loop left] node[left,text width=2.5cm]{\scriptsize
      (\texttt{FIN?~ACK!~ACK?})~or
      (\texttt{ACK?~FIN!})~or
      (\texttt{SIN!~ACK!})} (CLOSING);
		
		\draw[->] (CLOSING.south east) 
				  to
				  ([xshift=1.8cm]CLOSING.south east)
				  to
				  ([xshift=1.8cm,yshift=-0.4cm]CLOSING.south east)
				  arc(90:-90:0.15cm)
				  |-
				  ([xshift=1.8cm,yshift=-1.5cm]CLOSING.south east)
		          to
		          node[above right,text width=0.5cm]{\scriptsize \texttt{ACK?}} 
		          (TIME_WAIT);

		\draw[->] (ESTABLISHED) edge[right] node{\scriptsize \texttt{FIN!}} (FIN_WAIT_1);
		\draw[->] (ESTABLISHED) edge[below left] node[text width=0.5cm]{\scriptsize \texttt{FIN? ACK! ACK?}} (CLOSE_WAIT);
		\draw[->] (ESTABLISHED.south west) 
				to ([yshift=3cm]CLOSING)
				node[below left, text width=0.5cm]{\scriptsize \texttt{FIN? ACK! ACK?}}
		        to (CLOSING);

		\draw[->] (FIN_WAIT_1) edge[above left] node[text width=0.5cm]
		{\scriptsize \texttt{SYN! ACK!}} (CLOSING);
		\draw[->] (FIN_WAIT_1) 
		          to node[right]{\scriptsize \texttt{ACK?}} 
		          ([yshift=-1.3cm,xshift=0.6cm]FIN_WAIT_1)
		          arc(90:-90:0.15cm)
		          |-
		          (FIN_WAIT_2.north west);

		\draw[->] (FIN_WAIT_2) to node[left]{\scriptsize \texttt{FIN? ACK! ACK?}}
                              ([yshift=-0.9cm]FIN_WAIT_2)
                              arc(90:-90:0.15cm)
                              -| (TIME_WAIT.north);
		\draw[->] (FIN_WAIT_2) edge[above] node{\scriptsize \texttt{SYN! ACK!}} (CLOSING);

		\draw[->] (LAST_ACK) -| node[below right, text width=2cm]{\scriptsize (\texttt{SYN! ACK!}) or
      (\texttt{ACK? FIN!})} (CLOSING);
		\draw[->] (LAST_ACK) to node[above]{\scriptsize \texttt{ACK?}}
		                        ([xshift=1.8cm]LAST_ACK)
		                     |- (CLOSED);
		\draw[->] (LAST_ACK) to[loop below] node{\scriptsize \texttt{FIN? ACK! ACK?}} (LAST_ACK);

		\draw[->] (LISTEN)
		          to node[above]{\scriptsize \texttt{SYN!}}
              (SYN_SENT);
		\draw[->] (LISTEN.east) to 
		node[above,rotate=-15]{\scriptsize \texttt{ACK? FIN!}} (ESTABLISHED.north);

		\draw[->] (SYN_RECEIVED) edge[above] node[text width=2cm]{\scriptsize (\texttt{SYN!~ACK!})~or
      (\texttt{ACK?~FIN!})} (ESTABLISHED);
		\draw[->] (SYN_RECEIVED) to node[left,text width=1cm,xshift=0.3cm,yshift=-0.3cm]
		{\scriptsize \texttt{FIN? ACK! ACK?}}
		          ([yshift=-1cm]SYN_RECEIVED) to
		          ([yshift=-1cm,xshift=2.4cm]SYN_RECEIVED)
		          arc(180:0:0.15cm)
		          to ([yshift=-1cm,xshift=3.8cm]SYN_RECEIVED)
		          arc(180:0:0.15cm)
		          to ([yshift=-1cm,xshift=5.8cm]SYN_RECEIVED)
		          arc(180:0:0.15cm)
		          -| (CLOSE_WAIT.north east);
		\draw[->] (SYN_RECEIVED.west) to
		          ([xshift=-1cm]SYN_RECEIVED.west) to
		          node[right,rotate=90,xshift=-0.8cm,yshift=0.1cm]
		          {\scriptsize \texttt{FIN? ACK! ACK?}}
		          ([xshift=-1cm,yshift=-3cm]SYN_RECEIVED.south west) to
		          (CLOSING.north west);

		\draw[->] (SYN_SENT) edge[right] node[text width=2cm]
		          {\scriptsize (\texttt{SYN! ACK!}) or
		                       (\texttt{ACK? FIN!})}
		          (ESTABLISHED);

		\draw[->] (TIME_WAIT.north west) to node[below
    left]{\scriptsize{(\texttt{SYN! ACK!}) or (\texttt{ACK? FIN!})}}
		                                    ([yshift=-2cm]CLOSING.south west)
		                                 to (CLOSING.south west);
		\draw[->] (TIME_WAIT) edge[loop below] node{\scriptsize \texttt{FIN? ACK! ACK?}} (TIME_WAIT);

	\end{tikzpicture}}

%% file: fsm-figures/tcp/tcp.tex
\adjustbox{max width=\textwidth}{
	\centering
	\begin{tikzpicture}
		\node[state, initial] (CLOSED) {\scriptsize CLOSED};
		\node[state, below right=1cm and 0.5cm of CLOSED] (SYN_SENT) {\scriptsize SYN\_SENT};
		\node[state, below left=1cm and 0.5cm of CLOSED] (LISTEN) {\scriptsize LISTEN};
		\node[state, below=1cm and 0.5cm of LISTEN] (SYN_RECEIVED) {\scriptsize SYN\_RECEIVED};
		\node[state, below=1cm and 0.5cm of SYN_SENT] (ESTABLISHED) {\scriptsize ESTABLISHED};
		\node[state, below=1cm and 0.5cm of ESTABLISHED] (FIN_WAIT_1) {\scriptsize FIN\_WAIT\_1};
		\node[state, below right=1cm and 0.5cm of ESTABLISHED] (CLOSE_WAIT) {\scriptsize CLOSE\_WAIT};
		\node[state, below left=1cm and 0.5cm of FIN_WAIT_1] (CLOSING) {\scriptsize CLOSING};
		\node[state, below=0.5cm of FIN_WAIT_1] (FIN_WAIT_2) {\scriptsize FIN\_WAIT\_2};
		\node[state, below=1cm of CLOSE_WAIT] (LAST_ACK) {\scriptsize LAST\_ACK};
		\node[state, below=1cm of FIN_WAIT_2] (TIME_WAIT) {\scriptsize TIME\_WAIT};

		\draw[->] (CLOSED) edge[left] 
				  node{\scriptsize {\textcolor{blue}{\texttt{/* passive open */}}}} (LISTEN);
		\draw[->] (CLOSED) edge[above right, text width=2.2cm] 
				  node{\scriptsize {\textcolor{blue}{\texttt{/* active open */}}}
				                   \texttt{SYN!}} (SYN_SENT);
		\draw[->] (LISTEN) edge[left] node{\scriptsize
			\texttt{SYN? SYN! ACK!}} (SYN_RECEIVED);
		\draw[->] (LISTEN) edge[above, bend right] node[rotate=48]{\scriptsize \textit{timeout}} (CLOSED);
		\draw[->] (SYN_SENT) edge[right, text width=3.5cm] node{\scriptsize (\texttt{SYN? ACK?} or \texttt{ACK? SYN?}) \texttt{ACK!}} (ESTABLISHED);
		\draw[->] (SYN_SENT) edge[left] node{\scriptsize \texttt{SYN? ACK!}} (SYN_RECEIVED);
		\draw[->] (SYN_SENT) edge[above, bend left] node[rotate=-48]{\scriptsize \textit{timeout}} (CLOSED);
		\draw[->] (SYN_RECEIVED) edge[above] node{\scriptsize \texttt{ACK?}} (ESTABLISHED);
		\draw[->] (ESTABLISHED) edge[left] node{\scriptsize \texttt{FIN!}} (FIN_WAIT_1);
		\draw[->] (ESTABLISHED) edge[right] node{\scriptsize \texttt{FIN? ACK!}} (CLOSE_WAIT);
		\draw[->] (FIN_WAIT_1) edge[left] node{\scriptsize \texttt{FIN? ACK!}} (CLOSING);
		\draw[->] (FIN_WAIT_1) edge[right] node{\scriptsize \texttt{ACK?}} (FIN_WAIT_2);
		\draw[->] (CLOSE_WAIT) edge[right] node{\scriptsize \texttt{FIN!}} (LAST_ACK);
		\draw[->] (FIN_WAIT_2) edge[right] node{\scriptsize \texttt{FIN? ACK!}} (TIME_WAIT);
		\draw[->] (CLOSING) edge[below left] node{\scriptsize \texttt{ACK?}} (TIME_WAIT);
		\draw (LAST_ACK.south) edge[below,bend right] node[below]{\scriptsize \texttt{ACK?}} 
		      ([xshift=1.3cm]LAST_ACK.south);
		\draw[->] (5.5,0) -- (CLOSED);
		\draw (TIME_WAIT) -| (5.5,0);
	\end{tikzpicture}}
	\caption{\textbf{TCP Canonical FSM}.  User commands are shown in {\textcolor{blue}{\texttt{/* blue */}}}; you can view these as comments in the code (having no bearing on its logic).}
	\label{fig:tcp}

%% file: fsm-figures/dccp/dccp.tex
\adjustbox{max width=\textwidth}{
	\centering
	\begin{tikzpicture}

		\node[state, initial] (CLOSED) {\scriptsize CLOSED};
		\node[state, below left of=CLOSED] (LISTEN) {\scriptsize LISTEN};
		\node[state, below right of=CLOSED] (REQUEST) {\scriptsize REQUEST};
		\node[state, below=1cm and 0.5cm of LISTEN] (RESPOND) {\scriptsize RESPOND};
		\node[state, below=1cm and 0.5cm of REQUEST] (PARTOPEN) {\scriptsize PARTOPEN};
		\node[state, below=1cm and 0.5cm of RESPOND] (OPEN) {\scriptsize OPEN};
		\node[state, below=1cm and 0.5cm of PARTOPEN] (CLOSING) {\scriptsize CLOSING};
		\node[state, below of=OPEN] (CLOSEREQ) {\scriptsize CLOSEREQ};
		\node[state, below of=CLOSING] (TIMEWAIT) {\scriptsize TIMEWAIT};

		\draw[->] (CLOSED) edge[left] node{\scriptsize \texttt{active:=False}} (LISTEN);
		\draw[->] (CLOSED) edge[above right, text width=1.5cm] 
		               node[xshift=-0.2cm]{\scriptsize \texttt{active:=True; DCCP\_REQUEST!}} 
		      (REQUEST);

		\draw[->] (LISTEN) edge[right,text width=1.5cm] 
					   node{\scriptsize \texttt{DCCP\_REQUEST? DCCP\_RESPONSE!}} 
			  (RESPOND);

		\draw[->] (LISTEN) edge[below, bend right] node[rotate=55]{\scriptsize \textit{timeout}} 
		      (CLOSED);

		\draw[->] (REQUEST) edge[right,text width=1.5cm]
		                 node{\scriptsize \texttt{DCCP\_RESPONSE? DCCP\_ACK!}}
		      (PARTOPEN);

		\draw[->] (REQUEST.east) |- node[below right, text width=1.5cm]
		                     {\scriptsize \texttt{DCCP\_RESET?} or
		                      (\texttt{DCCP\_SYNC? DCCP\_RESET!}) or
		                      \textit{timeout}}
		      (CLOSED.east);

		\draw[->] (RESPOND) edge[left=3cm,text width=1.8cm]
		                node{\scriptsize \texttt{DCCP\_ACK?} or
		                                 \texttt{DCCP\_DATAACK?}}
		      (OPEN);

		\draw[->] (RESPOND) -| node[below, text width=3cm]
		                   {\scriptsize \textit{timeout}, then optionally
		                   \texttt{DCCP\_RESET!}}
		      (CLOSED);

		\draw[->] (RESPOND) edge[loop left] node{\scriptsize \texttt{DCCP\_DATA!}} (RESPOND);

		\draw[->] (PARTOPEN) edge[below,text width=5cm] 
		                 node[rotate=25,xshift=0.9cm]
		                 {\scriptsize (\texttt{DCCP\_DATA? DCCP\_ACK!}) or
		                                  (\texttt{DCCP\_DATAACK? DCCP\_ACK!}) or
		                                  \texttt{DCCP\_ACK?}} 
		      (OPEN); 

		\draw[->] (PARTOPEN) edge[loop right] node[rotate=60,xshift=-0.1cm,yshift=-0.2cm]
		{\scriptsize \texttt{DCCP\_DATAACK!}} (PARTOPEN);

		\draw[->] (PARTOPEN) edge[below] node[text width=2cm,rotate=-60,xshift=0.7cm]
						 {\scriptsize \textit{timeout} or
		                  (\texttt{DCCP\_CLOSE? DCCP\_RESET!})} 
		      (CLOSED);

		\draw[->] (PARTOPEN) edge[right, text width=1.8cm] 
		                 node{\scriptsize \texttt{DCCP\_CLOSEREQ? DCCP\_CLOSE!}} 
		      (CLOSING);

		\draw[->] (OPEN) edge[loop left,text width=1.9cm,below left] 
					 node{\scriptsize \texttt{DCCP\_DATA!}~or
                                      \texttt{DCCP\_DATAACK!}~or
                                      \texttt{DCCP\_ACK?}~or
                                      \texttt{DCCP\_DATA?}~or
                                      \texttt{DCCP\_DATAACK?}}
              (OPEN);

        \draw[->] (OPEN) edge[left,text width=1.9cm,left]
                     node[rotate=90,xshift=1cm,yshift=0.4cm]{\scriptsize \texttt{active==true;}
                                      \texttt{DCCP\_CLOSEREQ!}}
              (CLOSEREQ);

		\draw (OPEN.north west) -| (-6,-2);
		
		\draw[-stealth] (-6,-2) to
                        (CLOSED.south west);

        \draw[->] (OPEN) edge[bend right=35] node[below,text width=3cm]
                                    {\scriptsize (\texttt{active==true; DCCP\_CLOSE!})~or
                                                 (\texttt{DCCP\_CLOSEREQ? DCCP\_CLOSE!})}
                  (CLOSING);

        \draw (CLOSEREQ.west) -| node[below right,text width=3cm]
                                  {\scriptsize (\texttt{DCCP\_CLOSE? DCCP\_RESET!}) or,
		                               as en $\epsilon$-transition.}
                  (-6,-8);

        \draw[->] (CLOSING) to node[right] {\scriptsize \texttt{DCCP\_RESET?}} (TIMEWAIT);

        \draw (TIMEWAIT) to 
              ([yshift=-1.5cm]TIMEWAIT)
                         -| 
              (-6,-2);

        \draw (CLOSING.south west) to ([shift=({-4cm,-1.5cm})]TIMEWAIT);
	\end{tikzpicture}}
	\caption{\textbf{DCCP Canonical FSM}.}
	\label{fig:dccp}

%% file: synthesiserrors.tex

\subsection{Attack Synthesis Errors Examples}
Below we show examples of attacks that are
 synthesized with the canonical FSM, but not found with the NLP models. 

TCP.Canonical.3.9 spoofs both peers passively.  When tested against $\phi_3$, the attack causes the peers to end up in a deadlock in \texttt{SYN\_RECEIVED}$\times$\texttt{SYN\_RECEIVED}.  None of the TCP Gold, \linearrules, or \bertrules~attacks do passive spoofing; nor do any of them cause the peers to deadlock in \texttt{SYN\_RECEIVED}$\times$\texttt{SYN\_RECEIVED}.

DCCP.Canonical.2.18 spoofs both peers actively.  When tested against $\theta_2$, the attack causes the peers to navigate to \texttt{RESPOND}$\times$\texttt{RESPOND}.  On the way, they enter \texttt{TIME\_WAIT}$\times$\texttt{TIME\_WAIT}, violating $\theta_2$.  None of the DCCP Gold, \linearrules, or \bertrules~attacks do active spoofing; nor do any of them conclude in the state \texttt{RESPOND}$\times$\texttt{RESPOND}.  